\documentclass[twocolumn, preprintnumbers, amsmath, amssym]{revtex4-1}

\usepackage{graphicx}
\usepackage{dcolumn}
\usepackage{bm}
\usepackage{enumitem}
\usepackage{amssymb}
\usepackage{amsmath}
\usepackage{mathrsfs}
\usepackage{array}
\usepackage{float}
\usepackage{array}
\usepackage{multirow}
\usepackage{setspace}
\usepackage{tocloft}
\cftpagenumbersoff{table}

\begin{document}
\newcolumntype{C}[1]{>{\centering\arraybackslash}p{#1}}



\title{Evidence of a doublet wobbling excitation mode in  $^{105}$Pd}

\author{A. Karmakar$^{1,2}$ \altaffiliation[Present address: ]
{Tata Institute of Fundamental Research, Mumbai- 400005, India},\thanks, P. Datta$^{3}$,\thanks, N. Rather$^{4},$\thanks, S. Pal$^{5}$,\thanks, R. Palit$^{5}$,\thanks, A. Goswami$^{1,\#}$,\thanks, G. H. Bhat$^{6}$,\thanks, J. A. Sheikh$^{4,7}$, S. Jehangir$^{4}$}
\author{S. Chattopadhyay$^{1,2}$} 
\email{sukalyan.chattopadhyay@saha.ac.in}
\author{S. Frauendorf$^{8}$}

\affiliation{$^{1}$Saha Institute of Nuclear Physics, 1/AF, Bidhan Nagar, Kolkata - 700 064, India}
\affiliation{$^{2}$Homi Bhabha National Institute, Training School Complex, Anushakti Nagar, Mumbai - 400 094, India}
\affiliation{$^{3}$Ananda Mohan College, Kolkata - 700 009, India}
\affiliation{$^{4}$Department of Physics, Islamic University of Science and Technology, Awantipora, 192 122, India}
\affiliation{$^{5}$Tata Institute of Fundamental Research, Mumbai- 400005, India}
\affiliation{$^{6}$Department of Physics, GDC Shopian, Jammu and Kashmir, 192 303, India}
\affiliation{$^{7}$Department of Physics, University of Kashmir, Srinagar, 190 006, India}
\affiliation{$^{8}$Department of Physics, University of Notre Dame, Notre Dame, Indiana 46556, USA}
\affiliation{$^{\#}$Deceased}

\date{\today}

\begin{abstract}
An experimental investigation of $^{105}$Pd has revealed, for the first time, the existence of two wobbling bands, both having one phonon configuration and originating from excitation which is the wobbling from the yrast band with the $h_{11/2}$ quasineutron fully aligned with the short axis, and from an excited band with the same quasineutron but with less alignment along the short axis.  These observations have been drawn from the measured ratios of the inter-band and intra-band gamma transition rates. Model calculations based on the triaxial projected shell model (TPSM) approach have been performed and are found to be in good agreement with the experimental energies and relative transition probabilities. The analysis of the TPSM results provides an insight into the nature of the observed structures at a microscopic level.  
\end{abstract}

\maketitle
\section{Introduction}

The existence of triaxial deformed nuclei has been predicted by numerous mean-field studies (see e.g. Ref.~\cite{triaxial}) for an early systematic exploration of the nuclear chart). A significant feature of triaxial nuclei is the collective wobbling motion predicted by Bohr and Mottelson in 1975 \cite{bohr}. For large angular momenta, the energy spectrum will have a harmonic oscillator pattern \cite{bohr} with bands characterized by the signature quantum number that alternates with the number of a wobbling quantum. The nuclear wobbling is observed as consecutive rotational bands with increasing excitation energy, corresponding to successive wobbling quanta ($\propto n \hbar \omega_W$ for large I).  For these bands, the moments of inertia of the three principal axes are unequal and the total angular momentum wobbles about the medium (m) axis, which has the largest moment of inertia.

The wobbling mode, as originally predicted for an even-even system by Bohr and Mottelson, has not been observed yet. Instead, the first observation of the wobbling mode was reported in 2001 for the odd-mass $^{163}$Lu \cite{prl_hageman}. The authors interpreted their data in the framework of the particle + triaxial rotor (PTR) model (see e.g. Ref. \cite{bohr}). Two bands, labeled as TSD1 and TSD2, were identified, which were associated with the triaxial strongly deformed (TSD) minimum with odd-proton occupying the i$_{13/2}$ configuration, obtained in the ultimate cranking calculations of Ref. \cite{Bengtsson04}. The lower TSD1 band corresponds to $n$ = 0 and the excited TSD2 band to the $n=1$ wobbling excitation. The presence of the high-j proton significantly modifies the wobbling motion. Using the PTR model, Frauendorf and D\"onau \cite{dona} classified the collective mode into longitudinal wobbling (LW) when the total angular momentum vector (sum of particle and rotor angular momenta)  precesses about the medium axis of the triaxial nuclear shape and transverse wobbling (TW) when it precesses around an axis transverse to the medium axis. For increasing angular momentum, the excitation energy of the wobbling quanta increases for LW, and decreases for TW up to a critical value $I_c$, above which the TW changes into the LW regime. The wobbling mode represents a periodic motion of the three principal axes of the triaxial charge distribution, which generates strong E2 radiation. This is observed as collectively enhanced $\Delta I$ = 1 E2 transitions from the levels of the band with $n$ wobbling quanta to the band with $n-1$ wobbling quanta. Recently, the TW mode has been reported in several odd-A nuclei \cite{prl_hageman, garg, nandi, 127xe, 151eu, 163lu, 161lu, 165lu, 167lu, 167ta, 133la, 135pr, sensharmaa, sensharmaa2, 105pd_prl}.\par

The TW regime appears when the angular momentum $\mathbf{j}$ of the odd particle prefers to align with the short axis of the nucleus (or the odd hole with the long axis), which is transverse to the medium axis. 
In the yrast band of  $^{105}$Pd the odd quasineutron $h_{11/2}$ aligns with the
short axis with the maximal projection of $j_s=11/2$.
 There are excited bands that correspond to the reorientation of the odd particle's angular momentum $\mathbf{j}$, such that it precesses about the short axis with a projection $j_s<11/2$. Examples of such configurations with $j_s=9/2$  are the signature partner (SP) bands discussed in the framework of the PTR \cite{garg}.
 The name keeps the traditional nomenclature in high-spin physics, where the rotational bands group into $\Delta I =2$ signature sequences connected by strong intra-band E2 transitions. (The signature of the band is defined by $\alpha=I$ + even number.) In contrast to the TW  bands, SP bands are only connected by $\Delta I=1$ M1 transitions with the $n=0$ band. In all reported cases \cite{prl_hageman, garg, nandi, 127xe, 151eu, 163lu, 161lu, 165lu, 167lu, 167ta, 133la, 135pr, sensharmaa, sensharmaa2, 105pd_prl}, the SP band has been observed. Both the TW and signature partner bands have the signature, $-\alpha$, that is opposite to the signature  $\alpha$ of the $n = 0$ band.

\begin{figure}[!ht]
    \begin{center}
    \hspace*{-0.2cm}\includegraphics[width=8.5cm, angle =0]{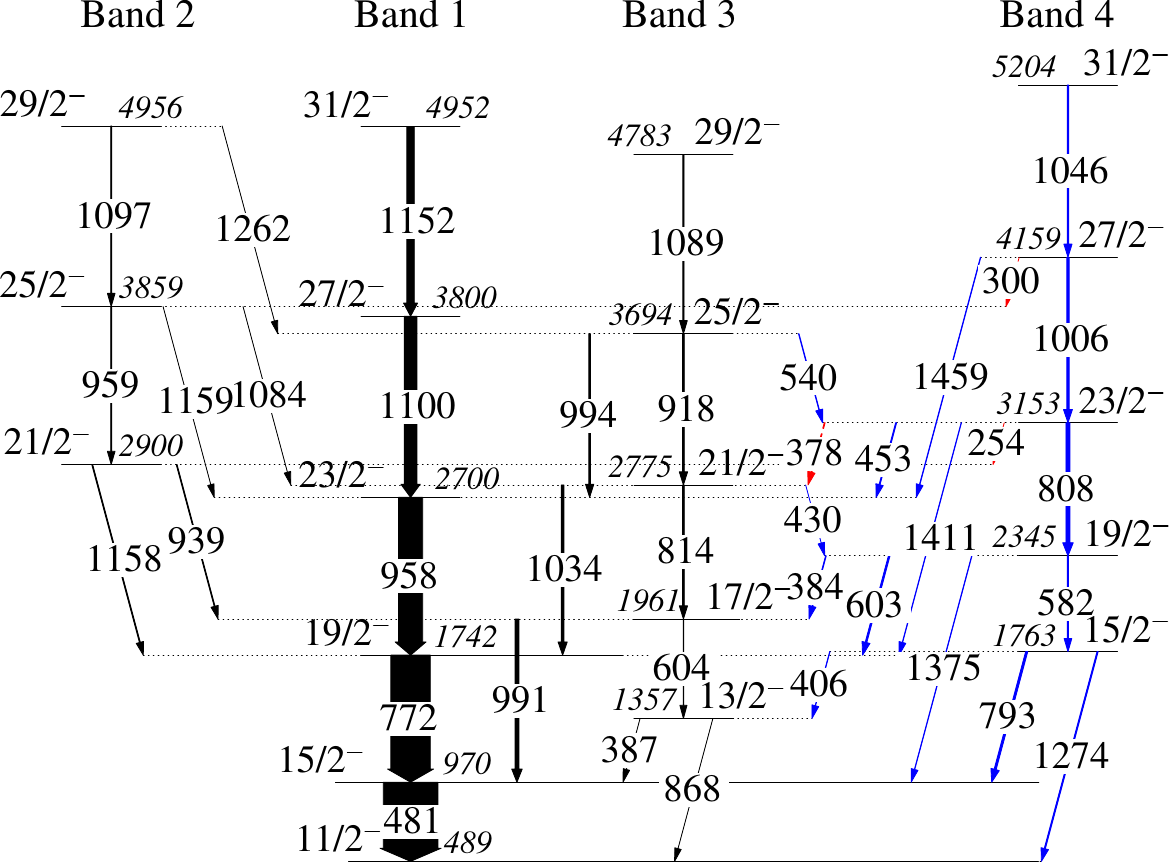}
    \end{center}
    \caption{The partial level scheme of $^{105}$Pd established from the present work. The level and transition energies are expressed in keV and have been rounded off to their nearest integer. The spins and energies of all the levels are with respect to the 11/2$^-$ isomeric level, which is 489.2 keV above the 5/2$^+$ ground state \cite{pd_old}. The levels and transitions marked in black were reported in \cite{105pd_prl} while those marked in blue were reported in \cite{pd_conf} by the same group. Our data confirms all these levels and transitions, and we have added three more transitions, which are marked in red. The spin and parities of Band 2 have been adapted from Ref.~\cite{105pd_prl}.}
\label{level_scheme_pd}
\end{figure}

In the case of $^{135}$Pr, the odd particle is a $h_{11/2}$ proton, and accordingly, the $n=0$ yrast band has $\alpha=-1/2$. The TW and SP bands with $\alpha=1/2$ were identified in Ref. \cite{garg}. The authors of Ref. \cite{sensharmaa} demonstrated that a fourth band with $\alpha=-1/2$ has the characteristics of the double TW excitation. The present case study of $^{105}$Pd has a structure analogous to $^{135}$Pr with the odd proton replaced by the odd $h_{11/2}$ neutron. The authors of Ref. \cite{105pd_prl} identified the $\alpha=-1/2$ yrast band (Band 1) and the $\alpha=1/2$ TW (Band 3) and SP (Band 2) bands depicted in Fig.~\ref{level_scheme_pd}. This work also observed, at a somewhat larger excitation energy, an $\alpha=-1/2$ band \cite{pd_conf} and the authors speculated that it might be the double TW band by analogy to $^{135}$Pr. 
However, they left the nature of the band an open problem because their data did not allow them to establish such an assignment.\par
Through detailed measurements of the mixing ratios of the transition connecting this band with the TW band, the present work provides experimental data for a different structural assignment. As will be demonstrated by means of comparing the data with the results 
of triaxial projected shell model (TPSM) calculations, the second $\alpha=-1/2$ band is not the double TW band as speculated, but it is a $n=0$ band with the reduced projection
$j_s=7/2$ and the second $\alpha=1/2$ band is the TW excited on it.\par
	\begin{figure*}[!ht]
		\begin{center}
			\hspace*{-0.2cm}\includegraphics[width=15cm, angle =0]{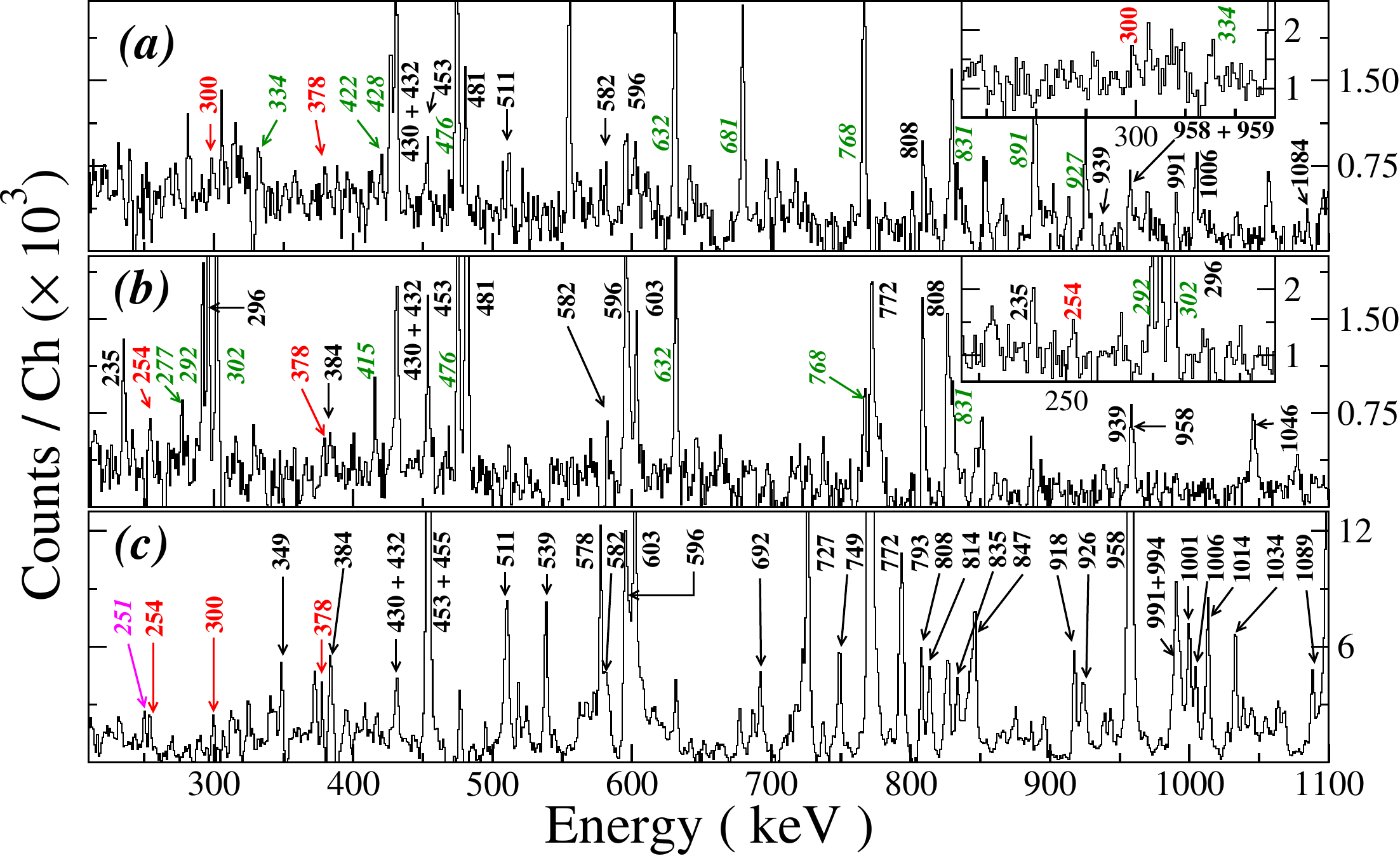}
		\end{center}
		\vspace{-0.5cm}
		\caption{The gamma spectra in coincidence with (a) 1046 keV (31/2$^-$ $\xrightarrow{}$ 27/2$^-$) keV, (b) 1006 (27/2$^-$ $\xrightarrow{}$ 23/2$^-$) keV transitions of Band 4 and (c) 481 (15/2$^-$ $\xrightarrow{}$ 11/2$^-$) transition of Band 1. The newly placed gamma transitions from Band 4 to Band 2 (300 and 254 keV) and Band 4 to Band 3 (378 keV) have been marked in red. The transitions marked in green and magenta belong to contamination from $^{102}$Ru and $^{104}$Pd, respectively. This contamination in (a) and (b) is due to the presence of 1047 keV (15$^-$ $\xrightarrow{}$ 13$^-$) and 1005 keV (14$^-$ $\xrightarrow{}$ 12$^-$) transitions in $^{102}$Ru. The insets of panels (a) and (b) show the sum gates of 1046 + 1158 keV and 1006 + 1158 keV, respectively.}
		\label{gated_pd}
	\end{figure*}
  
\section{Experimental Results}
	The high-spin states of $^{105}$Pd were populated through the fusion-evaporation reaction, where the 63 MeV $^{13}$C beam delivered by the 14-UD Pelletron at the Tata Institute of Fundamental Research (TIFR) was used to bombard a 1 mg/cm$^2$ enriched $^{96}$Zr target with Pb backing of 9 mg/cm$^2$ thickness. The deexciting $\gamma$ rays were detected using the Indian National Gamma Array (INGA) \cite{marcos}. During the time of the experiment, the array consisted of 18 Compton-suppressed clover detectors arranged in six rings with three at 40$^\circ$, two at 65$^\circ$, four at 90$^\circ$, three at 115$^\circ$, three at 140$^\circ$, and three at 157$^\circ$ with respect to the beam direction. The two and higher-fold coincidence data were recorded by a fast-digital data acquisition system based on Pixie-16 modules \cite{marcos}. The corresponding time-stamped data were sorted in a $\gamma$-$\gamma$ symmetric matrix using the multiparameter time-stamped based coincidence search (MARCOS) program, developed at TIFR \cite{marcos}. The matrix was used to establish the low-lying negative parity levels of $^{105}$Pd with the help of the RADWARE program LEVIT8R \cite{rad}. The photo-peak counts for the $\gamma$ rays in different gated spectra have been obtained by fitting the observed peaks to the Gaussian function using the INGASORT software \cite{iaeaINISRepository}. The uncertainty in the photo-peak count has been estimated in the present analysis by adding the statistical and fitting errors for the photo-peak, as these are correlated errors.\par

The partial level scheme of $^{105}$Pd is shown in Fig.~\ref{level_scheme_pd}, where the widths of the transitions are proportional to their relative intensities. Two new gamma transitions of 254 (23/2$^-$ $\xrightarrow{}$ 21/2$^-$) and 300 keV (27/2$^-$ $\xrightarrow{}$ 25/2$^-$) between Band 4 and Band 2 were observed in the 481 keV gated spectrum as shown in Fig.~\ref{gated_pd}(c). In order to ascertain the placement of these weak transitions, the gated spectra of 1046 keV (31/2$^-$ $\xrightarrow{}$ 27/2$^-$) keV and 1006 (27/2$^-$ $\xrightarrow{}$ 23/2$^-$) keV transitions are shown in Fig.~\ref{gated_pd}(a) and (b), respectively. These top gates show the presence of the 300 and 254 keV transitions, respectively, along with the following transitions of 1084 and 939 keV from Band 2 to Band 3. In order to further validate these placements, the 1158 keV gate has been added to these top gates, which corresponds to one of the feed-out transition energies of Band 2. These sum gates have been shown in the insets of Fig.~\ref{gated_pd}(a) and (b). It can be seen that the intensities of the 254 and 300 keV transitions exhibit a clear increase in these sum gates, thereby establishing that these transitions feed Band 2.\par

The relative intensities of the gamma transitions obtained from the present data agree well with the evaluated intensities from Ref.~\cite{pd_old}. However, the intensities of the newly observed gamma transitions in Ref.~\cite{105pd_prl} and \cite{pd_conf} were not reported. These have been evaluated from the present data and tabulated in Table~\ref{t1_pd}.\par

\begin{table*}[ht]
\caption{The energy (E$_\gamma$) and the relative intensity (I$_\gamma$) of the $\gamma$ rays of $^{105}$Pd along with the spin and parity of the initial (J$_i^\pi$) and the final (J$_f^\pi$) states, measured values of R$\mathrm{_{DCO}}$ and $\Delta\mathrm{_{iPDCO}}$ are shown. R$\mathrm{_{DCO}}$ values are obtained with gates on pure quadrupole transitions. The quoted uncertainties in the intensities include statistical and fitting errors only. The systematic error due to the efficiency determination has been estimated to be around 4\%.}

\small
\begin{tabular}{c c c c c c c c}
\hline
\label{t1_pd}

\centering

E$_\gamma$ & J$_i^\pi$ $\xrightarrow{}$ J$_f^\pi$ &I$_\gamma$ (rel.) & R$\mathrm{_{DCO}}$ & R$\mathrm{_{DCO}}$ &  P & $\delta$ & Mult. \\
(keV) & && (40$^\circ$) & (157$^\circ$) & & \\
\hline
\hline
 253.6(5) & 23/2$^-$ $\xrightarrow{}$ 21/2$^-$ &0.22(7) & 0.91(19) & 0.97(28) & - & 0.3$^{+0.2}_{-0.1}$ & Mixed\\
 299.6(5) & 27/2$^-$ $\xrightarrow{}$ 25/2$^-$ &0.12(5) & 0.89(21) & 0.85(31) & - & 0.3$^{+0.2}_{-0.1}$ & Mixed\\
 378.0(4) & 23/2$^-$ $\xrightarrow{}$ 21/2$^-$ &1.04(23) & 0.61(11) & - & - & - & M1\\
 383.9(4) & 19/2$^-$ $\xrightarrow{}$ 17/2$^-$ &0.84(21) & 0.66(12) & - & - & - & M1\\
 387.1(5) & 13/2$^-$ $\xrightarrow{}$ 15/2$^-$ &0.24(9) & - & - & - & - & -\\
 406.0(3) & 15/2$^-$ $\xrightarrow{}$ 13/2$^-$ &0.50(19) & - & - & - & - & -\\
 430.4(4) & 21/2$^-$ $\xrightarrow{}$ 19/2$^-$ &0.28(6) & 0.65(11) & - & - & - & M1\\
 453.4(5) & 23/2$^-$ $\xrightarrow{}$ 23/2$^-$ &1.19(25) & 0.82(15) & 0.83(18) & - & -0.6$^{+0.5}_{-0.6}$ & Mixed\\
 481.0(2) & 15/2$^-$ $\xrightarrow{}$ 11/2$^-$ &65.9(3) & 1.01(2) & & 0.57(3) & - & E2\\
 540.3(5) & 25/2$^-$ $\xrightarrow{}$ 23/2$^-$ &0.66(24) & 0.65(17) & - & - & - & M1\\
 581.9(3) & 19/2$^-$ $\xrightarrow{}$ 15/2$^-$ &1.38(37) & 1.06(15) & - & - & - & E2\\
 603.3(3) & 19/2$^-$ $\xrightarrow{}$ 19/2$^-$ &1.93(31) & 0.84(14) & 0.78(19) & -0.31(37) & -0.6$^{+0.3}_{-0.4}$ & Mixed\\
 604.0(5) & 17/2$^-$ $\xrightarrow{}$ 13/2$^-$ &0.64(11) & 0.97(15) & - & - & - & E2\\
 771.7(3) & 19/2$^-$ $\xrightarrow{}$ 15/2$^-$ &47.93(22) & 1.04(3) & - & 0.64(8) & - & E2\\
 793.1(4) & 15/2$^-$ $\xrightarrow{}$ 15/2$^-$ &2.48(28) & - & - & - & - & -\\
 808.4(4) & 23/2$^-$ $\xrightarrow{}$ 19/2$^-$ &4.81(39) & 0.96(10) & - & - & - & E2\\
 814.4(3) & 21/2$^-$ $\xrightarrow{}$ 17/2$^-$ &1.98(27) & 0.98(14) & - & - & - & E2\\
 868.1(4) & 13/2$^-$ $\xrightarrow{}$ 11/2$^-$ &0.42(11) & - & - & - & - & -\\
 918.3(3) & 25/2$^-$ $\xrightarrow{}$ 21/2$^-$ &2.18(29) & 1.00(12) & - & - & - & E2\\
 938.9(2) & 21/2$^-$ $\xrightarrow{}$ 17/2$^-$ &1.01(10) & 0.96(11) & - & - & - & E2\\
 958.3(4) &23/2$^-$ $\xrightarrow{}$ 19/2$^-$ & 29.51(29) & 0.98(2) & - & 0.46(14) & - & E2\\
 959.1(5) & 25/2$^-$ $\xrightarrow{}$ 21/2$^-$ &1.29(34) & - & - & - & - & -\\
 991.1(3) & 17/2$^-$ $\xrightarrow{}$ 15/2$^-$ &4.75(16) & 1.20(8) & 1.27(15) & -0.43(62) & 1.9$^{+0.4}_{-0.5}$ & Mixed\\
 993.8(5) & 25/2$^-$ $\xrightarrow{}$ 23/2$^-$ &2.16(15) & 1.08(10) & 1.23(14) & -0.36(49) & 2.4$^{+0.8}_{-0.5}$ & Mixed\\
 1005.6(4) & 27/2$^-$ $\xrightarrow{}$ 23/2$^-$ &2.90(48) & 1.05(12) & - & - & - & E2\\
 1033.8(3) & 21/2$^-$ $\xrightarrow{}$ 19/2$^-$ &3.11(16) & 1.13(8) & 1.15(14) & -0.42(41) & 2.2$^{+0.5}_{-0.4}$ & Mixed\\
 1045.6(5) & 31/2$^-$ $\xrightarrow{}$ 27/2$^-$ &1.93(37) & 1.03(13) & - & - & - & E2\\
 1083.7(4) & 25/2$^-$ $\xrightarrow{}$ 21/2$^-$ &0.48(16) & 1.03(21) & - & - & - & E2\\
 1089.3(5) & 29/2$^-$ $\xrightarrow{}$ 25/2$^-$ &1.38(32) & 1.05(19) & - & - & - & E2\\
 1097.0(3) & 29/2$^-$ $\xrightarrow{}$ 25/2$^-$ &1.25(33) & - & - & - & - & -\\
 1100.2(3) & 27/2$^-$ $\xrightarrow{}$ 23/2$^-$ &15.48(47) & 0.99(6) & - & 0.34(28) & - & E2\\
 1152.4(5) & 31/2$^-$ $\xrightarrow{}$ 27/2$^-$ &8.96(91) & 1.00(12) & - & 0.32(29) & - & E2\\
 1158.3(4) &21/2$^-$ $\xrightarrow{}$ 19/2$^-$ & 1.08(17) & - & 0.65(26) & - & - & M1\\
 1159.1(4) & 25/2$^-$ $\xrightarrow{}$ 23/2$^-$ &0.43(17) & - & - & - & - & -\\
 1262.3(3) & 29/2$^-$ $\xrightarrow{}$ 25/2$^-$ &0.18(9) & - & - & - & - & -\\
 1274.1(4) & 15/2$^-$ $\xrightarrow{}$ 11/2$^-$ &1.37(35) & - & - & - & - & -\\
 1375.0(2) & 19/2$^-$ $\xrightarrow{}$ 15/2$^-$ &0.51(34) & 1.09(24) & - & - & - & E2\\
 1411.4(3) & 23/2$^-$ $\xrightarrow{}$ 19/2$^-$ &0.51(32) & 0.98(23) & - & - & - & E2\\
 1458.7(4) & 27/2$^-$ $\xrightarrow{}$ 23/2$^-$ &0.60(35) & 0.98(21) & - & - & - & E2\\
\hline
\hline
\end{tabular}
\end{table*}

	\begin{figure}[!ht]
		\begin{center}
			\hspace*{-0.7cm}\includegraphics[width=8.1cm, angle =0]{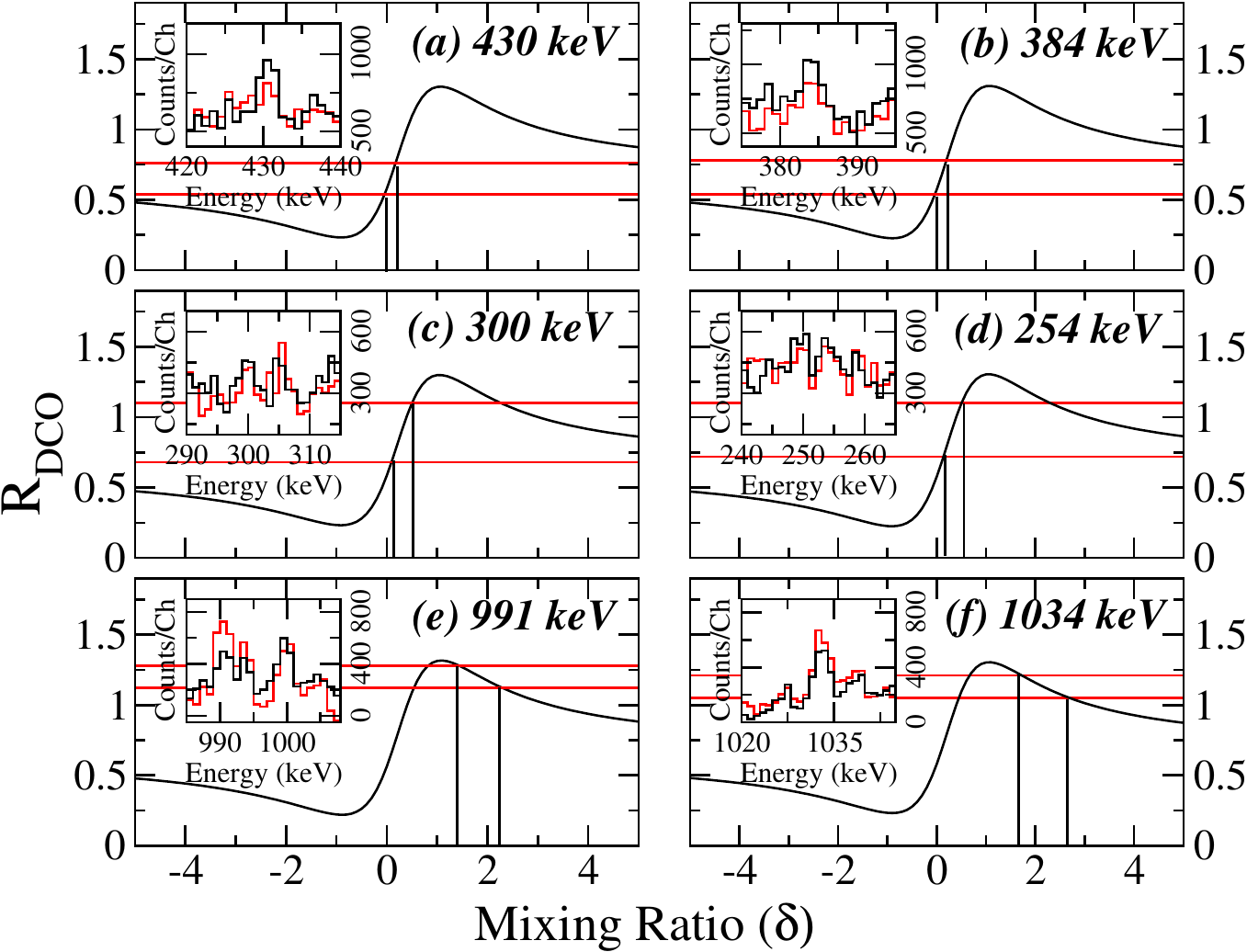}
		\end{center}
		\vspace{-0.5cm}
		\caption{Variation of the calculated R$\mathrm{_{DCO}}$ as a function of mixing ratio ($\delta$) for the $\Delta$I = 1 transitions between Band 3 and Band 4 ((a) 430 keV, (b) 384 keV), from Band 4 to Band 2 ((c) 300, (d) 254 keV) and from Band 3 to Band 1 ((e) 991 keV and (f) 1034 keV). The horizontal lines correspond to the R$\mathrm{_{DCO}}$ values at 40$^\circ$ along with their uncertainties. The corresponding gated spectra used for the R$_\mathrm{DCO}$ analysis are shown in the insets, in which the 90$^\circ$ and 40$^\circ$ spectra are marked with black and red colors, respectively.}
		\label{delta_dco_pd}
	\end{figure}
	\begin{figure}[!ht]
		\begin{center}
			\hspace*{-0.6cm}\includegraphics[width=8.5cm, angle =0]{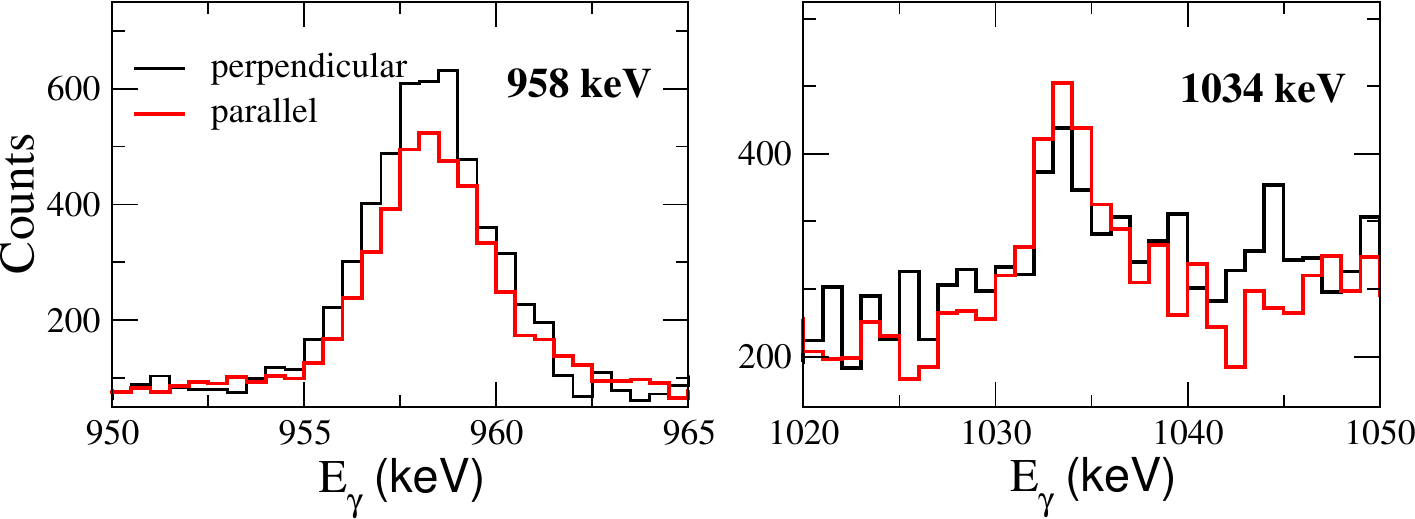}
		\end{center}
		\vspace{-0.5cm}
		\caption{Two gated $\gamma$-ray spectra used for the P measurements of 958 and 1034 keV transitions obtained from the sum gates of 481 and 772 keV transitions.}
		\label{pdco_pd}
	\end{figure}
	
	The $\gamma$ - ray multipolarities were determined from the Ratio of Directional Correlations from Oriented states (R$\mathrm{_{DCO}}$) method \cite{dco}. For this measurement, an asymmetric matrix was constructed with the $\gamma$ -ray energy detected at 90$^\circ$ along one axis while the coincident $\gamma$ -ray energy at 40$^\circ$ and 140$^\circ$ on the other axis. For the mixed transitions, the measured R$\mathrm{_{DCO}}$ values have been validated by re-measuring the value using the 157$^\circ$ and 90$^\circ$ matrix. For the R$\mathrm{_{DCO}}$ measurements, the stretched E2 gating transitions (481 and 772 keV) were used. The value for the attenuation coefficient of complete alignment ($\sigma$/J) of 0.31(3) was estimated from the measured R$\mathrm{_{DCO}}$ values for 1100 keV E2 and 1331 keV E1 transitions, which were assumed to be pure stretched transitions. The R$\mathrm{_{DCO}}$ values for 40$^\circ$ (157$^\circ$) were calculated using the computer code ANGCOR \cite{105pd_ang} (for $\sigma$/J = 0.31(3)) and were found to be 1.0 and 0.58 (1.0 and 0.51) for pure $\Delta$I = 2 and $\Delta$I = 1 transitions, respectively. As seen from Table~\ref{t1_pd}, the R$\mathrm{_{DCO}}$ values for all the in-band transitions of the four bands are within $\pm 1 \sigma$ for a pure $\Delta$I = 2 transition and thus, they have been assigned as E2 transitions. This is also consistent with previous assignments \cite{pd_old,105pd_prl}. Similarly, the DCO values for all the transitions between Bands 3 and 4 are consistent with pure $\Delta$I = 1 transitions, and the weighted mean of these values is 0.63(6). Thus, these transitions are predominantly M1 in nature. However, significant departures from the pure $\Delta$I = 1 values for the R$\mathrm{_{DCO}}$ ratios at 40$^0$ and 157$^0$ were observed for the transitions from Band 3 to 1 and Band 4 to 2 and also for the $\Delta$I = 0 transitions from Band 4 to 1. This difference in behaviors can be observed in the DCO spectra shown in the insets of Fig.~\ref{delta_dco_pd} (a), (b), (c), and (d), where 430 and 384 keV transitions are pure M1 transitions, while the other have mixed M1/E2 character.\par
    
\begin{figure}[!ht]
		\begin{center}
			\hspace*{-0.6cm}\includegraphics[width=8.5cm, angle =0]{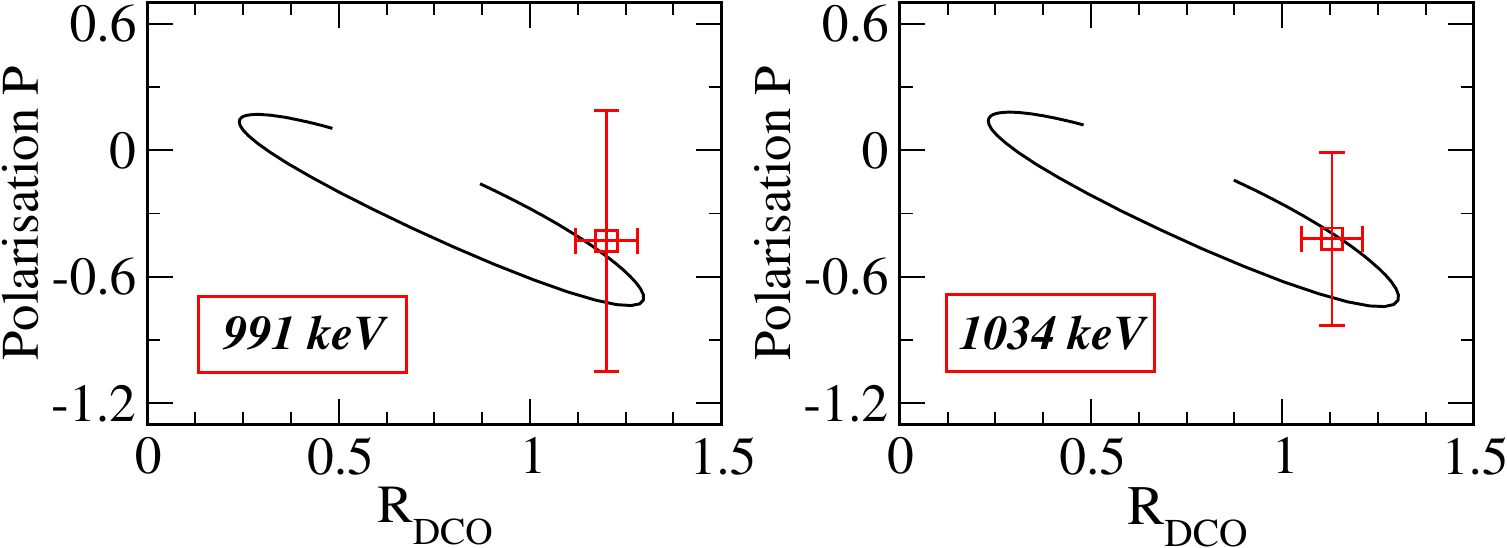}
		\end{center}
		\vspace{-0.5cm}
		\caption{Values of the DCO ratios (R$\mathrm{_{DCO}}$) and linear polarization (P) of the $\Delta I$ = 1 transitions that decay from the negative parity wobbling band to the positive parity ground state band, computed (solid line) and experimental (symbol) for various mixing ratios ($\delta$).}
		\label{contour_2_pd}
	\end{figure}

\begin{figure}[!ht]
		\begin{center}
			\hspace*{-0.6cm}\includegraphics[width=8.5cm, angle =0]{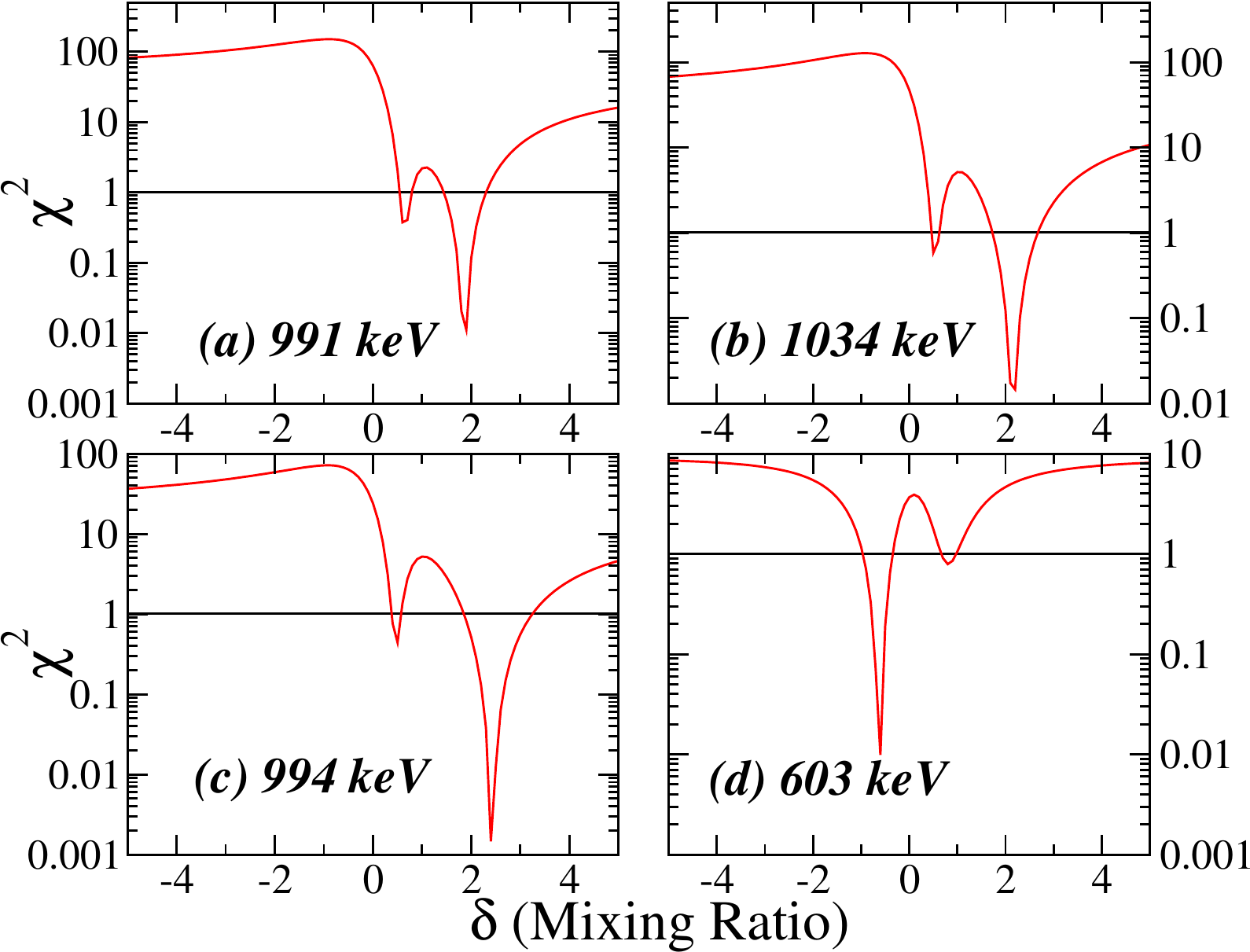}
		\end{center}
		\vspace{-0.5cm}
		\caption{The $\chi^2$ minimization plots for (a) 991 keV, (b) 1034 keV, (c) 994 keV, and (d) 603 keV. The horizontal lines correspond to the +1 of the minima value for determining the uncertainties. For the first three transitions, the lower minima correspond to the higher $\delta$ values. However, for 603 keV, the lower delta value is more probable.}
		\label{chi2_105pd}
	\end{figure}

For the unambiguous determination of the E2/M1 mixing ratio in a $\Delta$I = 1 transition, it is necessary to measure the linear polarization. Two representative $\gamma$-ray spectra for calculating the polarization asymmetry are shown in Fig.~\ref{pdco_pd}. The left panel shows a previously established electric transition, which has more counts in the perpendicular combination of crystals in 90$^\circ$ detectors than in the parallel combination. The opposite trend observed for the 1034 keV transition indicates the magnetic character of this transition. The mixing ratios for the 991, 1034, and 994 keV transitions between the wobbling bands were reported to be 1.8(5), 2.3(3), and 2.7(6), respectively \cite{105pd_prl}. The calculated values of the mixing ratio from polarization and R$\mathrm{_{DCO}}$ contour plots using the current experimental data for 991 keV and 1034 keV are shown in Fig.~\ref{contour_2_pd}. It may be observed from the figure that the large uncertainties on polarisation values do not allow the unique determination of the electromagnetic characters. However, the correlation between the P and R$_\mathrm{DCO}$ values in the contour plot can be used to obtain the lowest minimum from the $\chi^2$ minimization, thereby leading to the most probable value. This analysis has been shown in Fig.~\ref{chi2_105pd}, where the minima corresponding to the higher values of the mixing ratio for 991, 1034, and 994 keV transitions are the lowest. These values obtained from the $\chi ^2$ variation near the minimum are 1.9$^{+0.4}_{-0.5}$, 2.2$^{+0.5}_{-0.4}$ and 2.4$^{+0.8}_{-0.5}$, respectively, which are consistent with the previously reported values \cite{105pd_prl}. \par
	Alternately, the mixing ratio can be estimated by comparing the experimental  R$\mathrm{_{DCO}}$ value with the corresponding calculated values for the different values of $\delta$ using $\sigma$/J = 0.31(3) \cite{127xe}. These plots for the 991 and 1034 keV transitions are shown in Fig.~\ref{delta_dco_pd}(e) and (f), respectively. As can be seen in the figure, this analysis leads to two possible values of the mixing ratio, and only the higher values of the mixing ratio obtained from this method match the values obtained from the previous method within $\pm$1$\sigma$. Thus, this method has been followed to obtain the mixing ratios for low energy transitions of 254, 453 (E$_x$ = 3153 keV) and 300 keV (E$_x$ = 4159 keV), where the P measurements were not possible due to inadequate statistics. The results for 300 and 254 keV transitions are shown in Fig.~\ref{delta_dco_pd} (c) and (d), where the weighted mean value of the R$\mathrm{_{DCO}}$ ratio has been used. In this case also, two possible values of the mixing ratio are obtained. The higher $\delta$ value greater than 2.0 was found to be unrealistic as it leads to a much stronger (about 10 times) B(E2)$_{out}$ rate in I $\xrightarrow{}$ I$-$ 1 transitions than the B (E2)$_{in}$ rate in the I $\xrightarrow{}$ I$-$2 transitions. Thus, in the present analysis, we have estimated the mixing ratio values for the 254 and 300 keV transitions to be 0.3$^{+2}_{-1}$ and 0.3$^{+2}_{-1}$, respectively. It may be noted that although the error bars are large, the present statistics are adequate to establish significant E2 mixing in these low-energy $\Delta$I = 1 transitions. For the two $\Delta$I = 0 transitions of 453 and 603 keV from Band 4 to Band 3, the $\delta$ values were found to be $\approx$ -0.6(6) and -0.6(4), respectively. These values were reported as 0.0(7) and 0.0(5) by Rickey et al. \cite{pd_old}. It is not possible to compare the two measurements as the uncertainties reported are very large in both cases, and we may only conclude that the values match within $\pm$ 1$\sigma$. \par
	
    \begin{figure*}[!ht]
    \begin{center}
    \includegraphics[width=\linewidth]{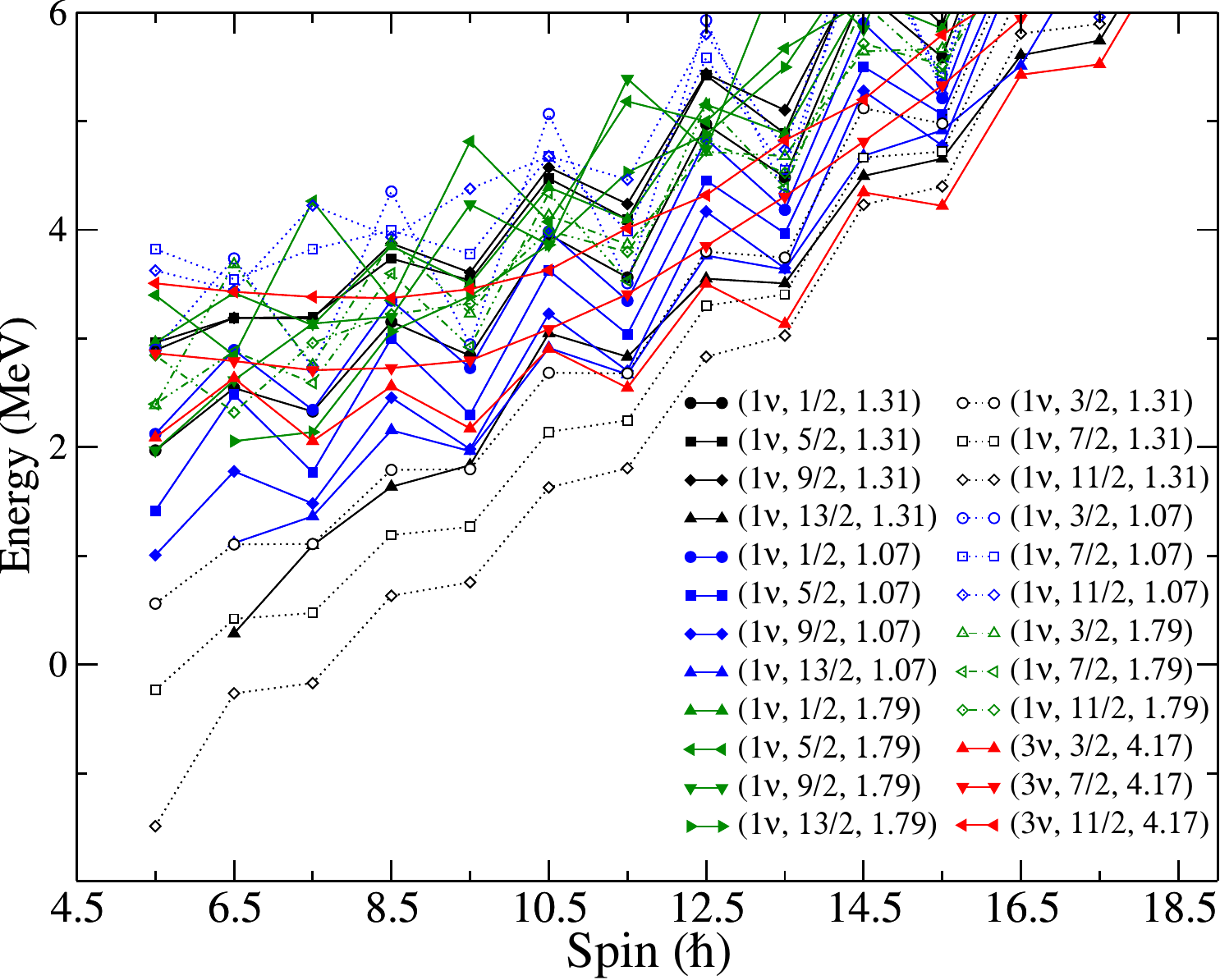}
    \end{center}
    \vspace{-0.5cm}
    \caption{Energies of the projected K configurations, where the short axis is chosen as the quantization axis 3 to which K refers. The curves are labeled by three quantities: quasiparticle character, K quantum number, and energy of the quasiparticle state. For instance, [1,11/2,1.31] designates the one quasineutron configuration projected on K= 11/2 having the BCS energy of 1.31 MeV.}
    \label{Band_before}
\end{figure*}
 \begin{figure}[t]
    \begin{center}
       \hspace*{-0.3cm}\includegraphics[width=5.5cm, angle =0]{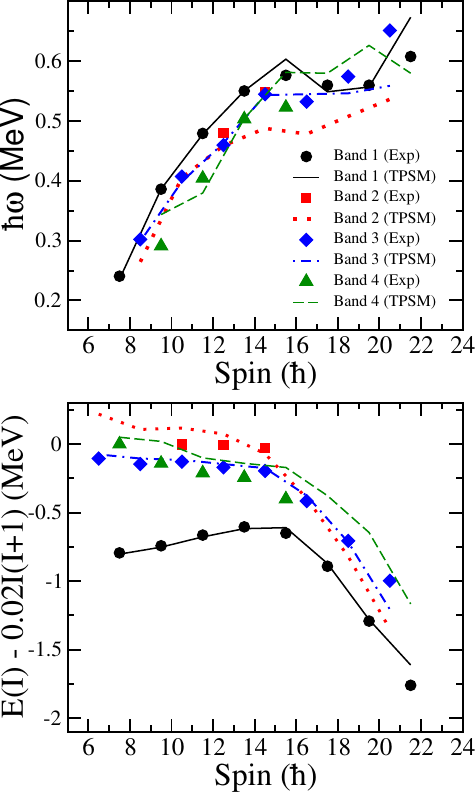}
    \end{center}
    \vspace{-0.5cm}
    \caption{Experimental and calculated values from TPSM for the rotational frequency (upper panel) and the level energies minus the rotor contribution (lower panel) as  functions of the spin I for Bands 1, 2, 3, and 4.}
    \label{energy}
\end{figure}
	The values of the ratio of the inter-band to intra-band transition rates have been listed in Table~\ref{t3_pd}, which shows two distinct classes of inter-band $\Delta$I = 1 transitions. In one case, the inter-band B(E2) values are of the same order as the intra-band B(E2) rates. Thus, these $\Delta$I = 1 transitions from Band 3 to Band 1 and Band 4 to Band 2 have substantial collective enhancement in the B(E2) rates thereby indicating the presence of the wobbling mode between these bands. On the other hand, the $\Delta$I = 1 transitions for Band 4 $\leftrightarrow$ Band 3 are predominantly M1 in nature, as the values of mixing ratio were found to be negligible. It is also apparent from Table~\ref{t3_pd} that the levels of Bands 3 and 4 exhibit an interesting duality. These levels decay by two $\Delta$I = 1 transitions, one with substantial E2 enhancement, while for the other, the E2 transition rate is negligible.
    
    \begin{figure*}[!ht]
    \begin{center}
    \includegraphics[width=\linewidth]{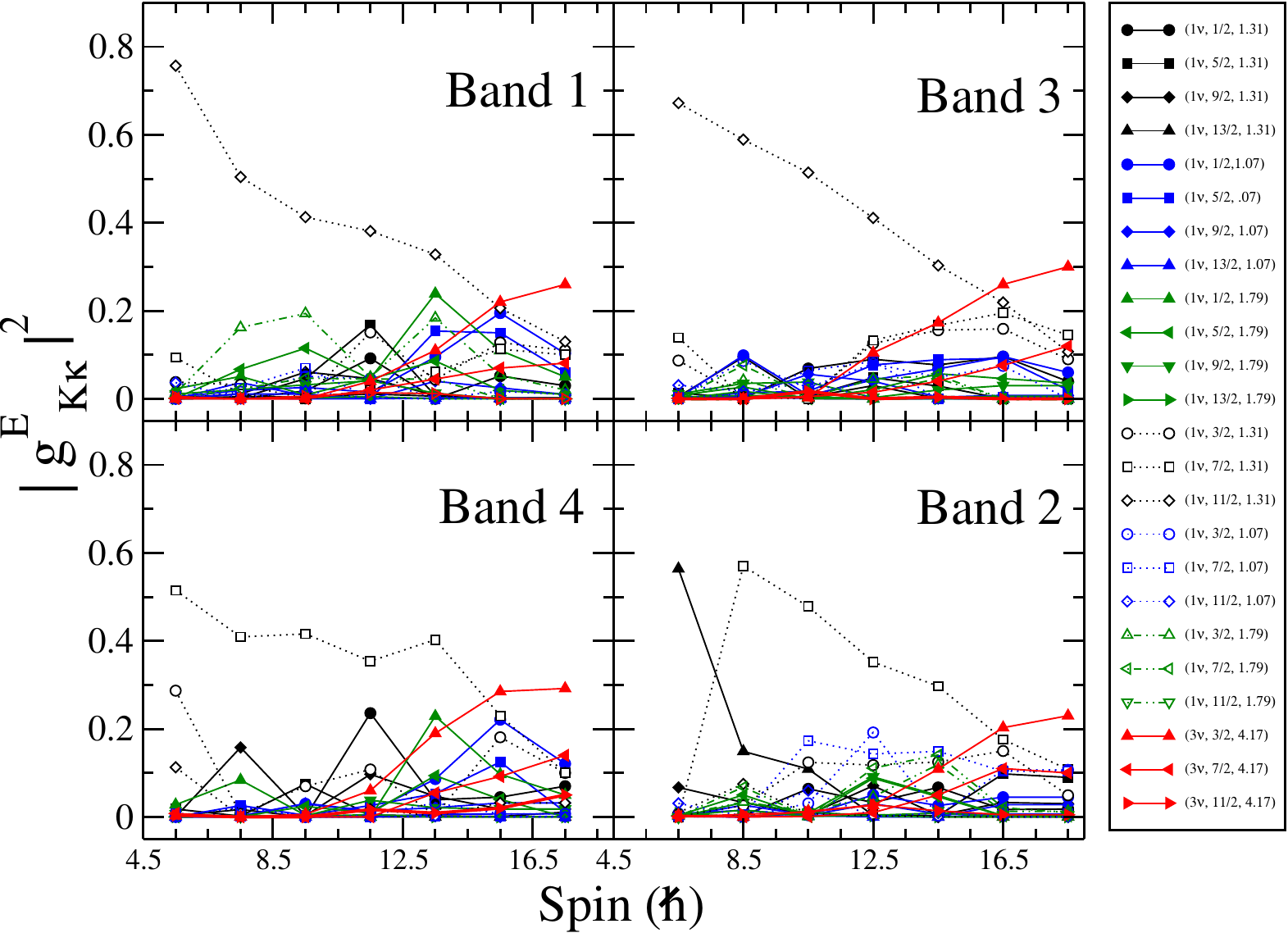}
    \end{center}
    \vspace{-0.5cm}
    \caption{\label{wave} Probabilities of various projected K configurations in the orthonormal basis shown in Fig.~\ref{Band_before}. The curves are labeled by three quantities: quasiparticle character, K quantum number, and energy of the quasiparticle state. For instance, [1,11/2,1.31] designates one quasineutron configuration projected on  K = 11/2 having the BCS energy of 1.31 MeV. The symbols and line types agree with Fig.~\ref{Band_before}.}
    \end{figure*}

\begin{table*}[!ht]	
\caption{The energy (E$_\gamma$) and the mixing ratios ($\delta$) of the $\Delta$I = 1 $\gamma$ rays of $^{105}$Pd, along with the spin and parity of the initial (J$_i^\pi$) and the final (J$_f^\pi$) states, experimentally and calculated values of B(E2)$\mathrm{_{out}}$/B(E2)$\mathrm{_{in}}$ and B(M1)$\mathrm{_{out}}$/B(E2)$\mathrm{_{in}}$ are tabulated. For mixing ratio values of 0.1 or less, the effect of mixing ratio on B(M1)$\mathrm{_{out}}$/B(E2)$\mathrm{_{in}}$ is found to be negligible. Hence, only the upper limits of the B(E2)$\mathrm{_{out}}$/B(E2)$\mathrm{_{in}}$ ratios have been tabulated. The value of the DCO ratio for the 1159.1 keV transition of Band 2 could not be estimated. The tabulated value was calculated by assuming a pure M1 character for this transition.}

\small

\begin{tabular}{c c c c c c c}
\hline
\label{t3_pd}

\centering
& & &  Band 2 & & & \\
\hline
 J$_i^\pi$ $\xrightarrow{}$ J$_f^\pi$ &E$_\gamma$ & $\delta$ & \multicolumn{2}{c}{B(E2)$\mathrm{_{out}}$/B(E2)$\mathrm{_{in}}$} & \multicolumn{2}{c}{B(M1)$\mathrm{_{out}}$/B(E2)$\mathrm{_{in}}$} \\
(keV) & & & Expt. & Calc. & Expt. & Calc. \\
\hline
  25/2$^-$ $\xrightarrow{}$ 23/2$^-$ &1159.1(4)& - & - & - & 0.16(8) & 0.22\\
\hline
\\
& & &  Band 3 & & & \\
\hline
J$_i^\pi$ $\xrightarrow{}$ J$_f^\pi$ &E$_\gamma$ & $\delta$ & \multicolumn{2}{c}{B(E2)$\mathrm{_{out}}$/B(E2)$\mathrm{_{in}}$} & \multicolumn{2}{c}{B(M1)$\mathrm{_{out}}$/B(E2)$\mathrm{_{in}}$} \\
(keV) & & & Expt. & Calc. & Expt. & Calc. \\
\hline
 17/2$^-$ $\xrightarrow{}$ 15/2$^-$ &991.1(3) & 1.9$^{+0.4}_{-0.5}$ & 0.49(15) & 0.46 & 0.09(3) & 0.19\\
 21/2$^-$ $\xrightarrow{}$ 19/2$^-$  &430.4(4) & 0.1 $\downarrow$ & 0.03 $\downarrow$ & 0.01 & 0.44(13) & 0.58\\
 & 1033.8(3) &  2.2$^{+0.5}_{-0.4}$ & 0.39(10) & 0.40 & 0.06(2) & 0.15\\
 25/2$^-$ $\xrightarrow{}$ 23/2$^-$ & 540.3(5) &0.1 $\downarrow$ & 0.04 $\downarrow$ & 0.001 & 0.86(35) & 0.64\\
 & 993.8(5) & 2.4$^{+0.8}_{-0.5}$ & 0.57(17) & 0.32 & 0.07(2) & 0.15\\
 
\hline
\\
& & &  Band 4 & & & \\
\hline
J$_i^\pi$ $\xrightarrow{}$ J$_f^\pi$ &E$_\gamma$ & $\delta$ & \multicolumn{2}{c}{B(E2)$\mathrm{_{out}}$/B(E2)$\mathrm{_{in}}$} & \multicolumn{2}{c}{B(M1)$\mathrm{_{out}}$/B(E2)$\mathrm{_{in}}$} \\
(keV) & & & Expt. & Calc. & Expt. & Calc. \\
\hline
19/2$^-$ $\xrightarrow{}$ 17/2$^-$  & 383.9(4) &  0.1 $\downarrow$ & 0.05 $\downarrow$ & 0.01 & 0.50(16) & 0.61 \\
23/2$^-$ $\xrightarrow{}$ 21/2$^-$  & 253.6(5) & 0.3$^{+0.2}_{-0.1}$ & 1.23(75) & 0.83 & 0.61(37) & 0.51\\
& 378.0(4) & 0.1 $\downarrow$ & 0.09 $\downarrow$ & 0.01 & 0.96(24) & 0.53\\
27/2$^-$ $\xrightarrow{}$ 25/2$^-$  & 299.6(5) & 0.3$^{+0.2}_{-0.1}$ & 1.45(98) & 0.74 & 1.01(68) & 0.51\\

\hline

\hline
\hline
\end{tabular}
\end{table*}

\begin{figure}[t]
    \begin{center}
    \includegraphics[width=0.9\linewidth]{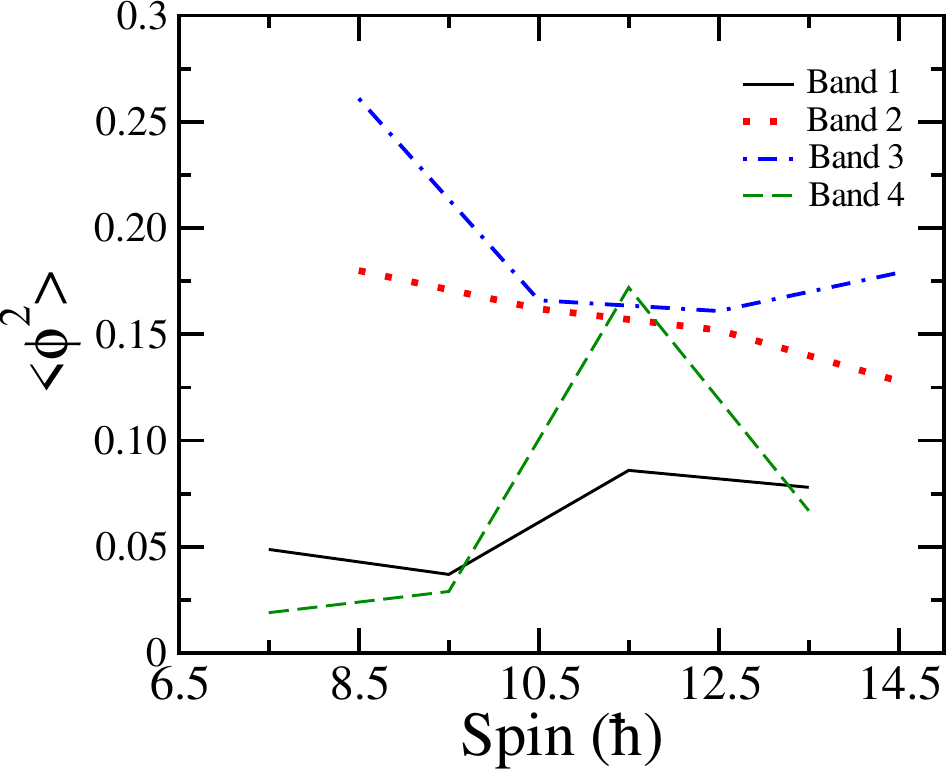}
    \end{center}
    \caption{\label{delta-phi} Mean value $\langle\phi^2\rangle$ of the angular momentum $\bm{J}$ angle $\phi$ with the short axis within the short-medium plane.}
    \end{figure}

	\begin{table}[!ht]
		\caption{Transition rates of different bands from TPSM Calculations used for the comparison plots with experimental values.}
		\label{t2_pd}
		\centering
		\begin{tabular}{c c c c c}
			& &\hspace{-0.5cm}Band 2 & &\\
			\hline
			Spin & B(E2)$_{in}$  & B(E2)$_{out}$  & B(M1)$_{out}$ & Connected to\\
			($\hbar$) & (e$^2$b$^2$)  & (e$^2$b$^2$)   & ($\mu_n^2$) & \\
			\hline
			\hline
			10.5 & 0.54 & 0.008 & 0.13 & Band 1 ($\Delta$I = 1)\\
			& &  0.11 & & Band 3 ($\Delta$I = 2)\\
            & &  0.22 & 0.14& Band 3 ($\Delta$I = 0)\\
             &  & 0.335 & 0.0003 & Band 4 ($\Delta$I = 1)\\
			12.5 & 0.82 & 0.008 & 0.18 & Band 1 ($\Delta$I = 1)\\
			& &  0.15 & & Band 3 ($\Delta$I = 2)\\
            & &  0.24 & 0.39 & Band 3 ($\Delta$I = 0)\\
             &  & 0.338 & 0.0009 & Band 4 ($\Delta$I = 1)\\
			14.5 & 0.89 & 0.001 & 0.19 & Band 1 ($\Delta$I = 1)\\
            & &  0.09 & & Band 3 ($\Delta$I = 2)\\
            & &  0.25 & 0.41 & Band 3 ($\Delta$I = 0)\\
            &  & 0.369 & 0.0002 & Band 4 ($\Delta$I = 1)\\
			\hline
			& & & & \\
		\end{tabular}
		
		\begin{tabular}{c c c c c}
			& & \hspace{-0.5cm}Band 3 & &\\
			\hline
			Spin & B(E2)$_{in}$  & B(E2)$_{out}$  & B(M1)$_{out}$ & Connected to\\
			($\hbar$) & (e$^2$b$^2$)  & (e$^2$b$^2$)   & ($\mu_n^2$) & \\
			\hline
			\hline
			8.5 & 0.72 & 0.33 & 0.14 & Band 1 ($\Delta$I = 1)\\
			& & 0.01 & 0.41 & Band 4 ($\Delta$I = 1)\\
			10.5 & 0.80 & 0.32 & 0.12 &  Band 1 $\Delta$I = 1)\\
			& & 0.01 & 0.46 & Band 4 ($\Delta$I = 1)\\
			12.5 & 0.72 & 0.23 & 0.11 & Band 1  ($\Delta$I = 1)\\
			& & 0.001 & 0.46 & Band 4 ($\Delta$I = 1)\\
			14.5 & 0.70 & 0.18 & 0.11 &  Band 1 ($\Delta$I = 1)\\
			\hline
			& & & & \\
		\end{tabular}
		
		\begin{tabular}{c c c c c}
			& & & \hspace{-0.5cm}Band 4  &\\
			\hline
			Spin & B(E2)$_{in}$  & B(E2)$_{out}$  & B(M1)$_{out}$ & Connected to\\
			($\hbar$) & (e$^2$b$^2$)  & (e$^2$b$^2$)   & ($\mu_n^2$) & \\
			\hline
			\hline
			7.5 & 0.63 & 0.02 & 0.51 & Band 3 ($\Delta$I = 1)\\
            & & 0.58 & 0.36 & Band 2 ($\Delta$I = 1)\\
			& & 0.23 & 0.08 & Band 1 ($\Delta$I = 0)\\
			& & 0.02 &  & Band 1 ($\Delta$I = 2)\\
			9.5 & 0.69 & 0.01 & 0.42 & Band 3 ($\Delta$I = 1)\\
			& & 0.53 & 0.35 & Band 2 ($\Delta$I = 1)\\
			& & 0.20 & 0.17 & Band 1 ($\Delta$I = 0)\\
			& & 0.02 &  & Band 1 ($\Delta$I = 2)\\
			11.5 & 0.72 & 0.01 & 0.38 & Band 3 ($\Delta$I = 1)\\
			& & 0.60 & 0.37 & Band 2 ($\Delta$I = 1)\\
			& & 0.28 & 0.79 & Band 1 ($\Delta$I = 0)\\
			& & 0.01 &  & Band 1 ($\Delta$I = 2)\\
			13.5 & 0.74 & 0.004 & 0.35 & Band 3 ($\Delta$I = 1)\\
			& & 0.55 & 0.38 & Band 2 ($\Delta$I = 1)\\
			& & 0.27 & 0.65 & Band 1 ($\Delta$I = 0)\\
			& & 0.01 &  & Band 1 ($\Delta$I = 2)\\
			\hline
		\end{tabular}
	\end{table}
	
\section{Discussion}
To provide an insight into the nature of the band structures observed in $^{105}$Pd, TPSM \cite{rev_tpsm,tpsm,rev_tpsm_tpsm} calculations have been performed. This model has been shown to provide a unified description of the interplay between collective and single-particle degrees of freedom in deformed odd-mass nuclei \cite{SJ21, SB10, JM17, 151eu}. The model is based on an augmented version of the pairing plus
quadrupole-quadrupole Hamiltonian of Baranger and Kumar 
\cite{https://doi.org/10.1016/0375-9474(68)90044-4}. The Hamiltonian of the TPSM approach is given by 
\begin{equation}
\hat H =  \hat H_0 -  {1 \over 2} \chi \sum_\mu \hat Q^\dagger_\mu
\hat Q^{}_\mu - G_M \hat P^\dagger \hat P   - G_Q \sum_\mu \hat
P^\dagger_\mu\hat P^{}_\mu , \label{hamham}
\end{equation}
where $\hat H_0$ is the spherical part of the Nilsson potential
\begin{equation}
\hat H_N = \hat H_0 - {2 \over 3}\hbar\omega\left\{\epsilon\hat Q_0
+\epsilon'{{\hat Q_{+2}+\hat Q_{-2}}\over\sqrt{2}}\right\}.
\label{nilsson}
\end{equation}
The axial deformation parameter $\varepsilon$ and
triaxiality parameter $\varepsilon'$ (or $\gamma=\arctan(\varepsilon'/\varepsilon$)) are
 input parameters of the model, for which we choose $\epsilon$ = 0.26 and $\epsilon'$ = 0.11 ($\gamma=23^\circ$) in accordance with the particle + triaxial rotor calculations in Ref. \cite{105pd_prl}.  
The Nilsson Hamiltonian is diagonalized, providing the triaxial single-particle states. The monopole  pairing 
is treated by the standard Bardeen–Cooper–Schrieffer (BCS) approximation based 
on the  Nilsson energies.
The monopole pair strength  is of the standard form \cite{45jm}
\begin{eqnarray}
G_M = {{G_1 \mp G_2{{N-Z}\over A}}\over A}, 
 \label{pairing}
\end{eqnarray}
where the minus (plus) sign applies to neutrons (protons). 
In the present calculation, we have chosen $G_1=20.12$ and $G_2=13.13$, which approximately reproduce the observed odd-even mass difference in the mass region. 
The quadrupole pairing strength $G_Q=0.18G_M$. 
The quadrupole coupling constant $\chi$
is determined via self-consistency by the input parameter $\varepsilon$
(see details e. g. in Ref. \cite{Rouoof25}).

Nilsson + BCS multi quasiparticle configurations $\vert\kappa\rangle$ are then constructed, where $\kappa$ indicates the number of quasiparticles and the BCS energy of the configuration. For the odd N nucleus $^{105}$ Pd, the configurations of one and three quasiparticles in the neutron sector and the configurations of zero and two quasiparticles in the proton sector are taken into account. 
These intrinsic states are projected onto good angular-momentum using the projection operator $P^I_{MK}$ \cite{ring80,Hara332,Hara348}, which generates the TPSM basis $P^I_{MK}\vert\kappa\rangle$.
These basis states are used to diagonalize the TPSM  Hamiltonian (\ref{hamham}). 

The energy expectation values of the projected states, 
$\langle\kappa\vert P^I_{KM} HP^I_{MK}\vert\kappa\rangle/
\langle\kappa\vert P^I_{KM} P^I_{MK}\vert\kappa\rangle$ (independent of $M$) are shown in Fig.~\ref{Band_before}, what is called as ``band diagram". The basis states are arranged in adiabatic rotational bands with a fixed angular momentum projection $K$ on the {\it short} principal axis of the triaxial body. The band diagram provides information on the intrinsic structure of the bands before diagonalization of the TPSM Hamiltonian (\ref{hamham}).

It is noted that the axis of choice is different from the TPSM calculations published so far, where $K$ refers to the long principal axis. The present choice of short axis simplifies the interpretation, because 
the odd quasineutron tends to align its angular momentum with the short axis. This has the consequence that the final states are much less mixed with respect to $K$. Of course, such a change of the quantization axis leaves the observables invariant. Appendix \ref{secA} explains how the change of the quantization axis is achieved by changing the $\gamma$ degree of freedom.








The original TPSM  basis set  $P^I_{MK}\vert\kappa\rangle$ is non-orthogonal. 
For the interpretation, it is common to change to the orthonormal basis $\vert IMK\kappa\rangle$ such that the TPSM eigenstates \cite{tpsm,wang20,nazir23}
are 
\begin{align}
    \vert IME\rangle=\sum_{K\kappa'} g_{K\kappa}^{E}\vert IMK\kappa\rangle.
    \label{eq:ortho}
\end{align} 
For details of the above procedure, the reader is referred to Ref.~\cite{rev_tpsm}.

Fig.~\ref{energy} compares the experimental energies $E(I)$ and the derived rotational frequencies  $\hbar\omega(I)$  with the TPSM results. The TPSM values reproduce the experimental quantities for all
four bands quite well. The leveling of Band 1 at $\hbar\omega$ = 0.58 MeV (``up bend'' of the function $I(\omega)$) reflects
the rotational alignment of a pair of $h_{11/2}$ quasineutrons, which is seen in Fig.~\ref{Band_before} as the bands originating from the three-quasineutron configuration at 4.17 MeV becoming yrast. A similar leveling is found in the TPSM results for Bands 2, 3, and 4, which reflects the crossing of the three-quasineutron configurations. It is seen in the experimental frequencies of Bands 3 and 4 as well. The TPSM results are consistent with Ref.~\cite{105pd_prl}, which assigned the leveling to the alignment of two quasineutrons as well.

In the discussion of the wobbling nature of the band structures, it is important to ascertain the decomposition of the wavefunction in terms of the
basis configurations.
Fig.~\ref{wave} displays the probabilities $\vert g_{K\kappa}^{E}\vert ^2$ in the orthonormal basis (\ref{eq:ortho}).
The labels $(K,\kappa)$ in the orthonormal basis have a similar but not the same meaning as the original labels, ($K,\kappa$), because the orthonormalization mixes the projected quasiparticle configurations. 


The lowest sequences of basis states in Fig.~\ref{Band_before}  are projections from the $h_{11/2}$ one-quasineutron configuration at 1.31 MeV onto different values of $K$. As seen in Fig.~\ref{wave}, for the states of $I$ = 11/2, 15/2 19/2, ... of  Band 1, the largest components
are from the $K$ = 11/2 states.  For the $I$ = 13/2, 17/2, 21/2, ... states of Band 3, the same $K$ = 11/2 basis states appear with the largest probability. Decomposing these basis states into their quasineutron factors and collective factors (defined as the projections from the quasiparticle vacuum) shows that the main quasineutron component is $h_{11/2}$ with the projection of the triaxial Nilsson state along the s-axis, $k$ = 11/2. The higher excitation energies of the $I$ = 13/2, 17/2, 21/2, ... states compared to the
$I$ = 11/2, 15/2, 19/2....states, as seen in Fig.~\ref{Band_before}, are due to a larger contribution of the collective angular momentum. This is the expected structure for the wobbling mode with the angular momentum of the $h_{11/2}$ quasineutron being aligned with the s-axis. Fig. \ref{delta-phi} shows that the average deviation of its total angular momentum from the s-axis is much larger for Band 3 than for Band 1. This indicates a collective wobbling excitation mode.  Exciting the $n=1$ TW state seems not to strongly change its quasineutron component.  

For the $I$ = 15/2, 19/2, 23/2, ... states of Band 4, the largest component is $K$ = 7/2,  and decomposing it into the quasineutron and collective factors shows that the most probable component is  $h_{11/2}$ with the $k =$ 7/2 projection.  It has a larger energy than the $K$ = 11/2 component of Band 1 in Fig.~\ref{Band_before}. Accordingly, Band 4 lies above Band 1. For $I$ = 17/2, 21/2, .... of Band 2, the dominating basis state is $K$ = 7/2 as well. Fig. \ref{delta-phi} shows that the average deviation of its total angular momentum from the s-axis is much larger than for Band 4. This indicates a collective wobbling excitation. We interpret Band 2 as a collective wobbling excitation based on Band 4 because it emerges from the same $k=7/2$ quasineutron configuration. Because of its collective wobbling energy, it lies above Band 4.

Fig.~\ref{delta-phi} provides the essential evidence of the above interpretation. It displays the mean value $\langle\phi^2\rangle$, where $\phi$ is the angle of the total angular momentum vector $\bm{J}$ with the short axis within the short-medium plane of the triaxial shape. The angle $\phi$ is given by the Fourier transform of the reduced density matrix $\rho(K, K')=\sum_\kappa g^{E}_{K,\kappa}g^{E}_{K',\kappa}$. The details can be found in Ref. \cite{SSS}. (Note, $K$ refers to projection along the long axis there. The $g^{E}_{K,\kappa}$ used in Fig.~\ref{delta-phi} are calculated for $\gamma=23^\circ$.) The small values for Bands 1 and 4 indicate that these represent uniform rotation about the short axis of the two different $h_{11/2}$ quasineutron configurations with the main components $k=11/2$ and 7/2. The large values for Bands 3 and 2 indicate that $\bm{J}$ moves away from the short axis i.e., it wobbles. Naturally, Band 3 is the wobbling excitation of Band 1, and Band 2 is the wobbling excitation of Band 4, as they are based on the same respective quasineutron structures. As discussed in Ref. \cite{SCS, SSS}, the transverse geometry becomes unstable at a critical value $I_c$ where $\mathbf{J}$ moves far into the s-m plane. The large value for $I=21/2$ of Band 4 may indicate the instability of the transverse geometry. Its return to the smaller value at higher $I$ may be due to the alignment of the pair of $h_{11/2}$ quasineutrons, which stabilize the transverse geometry.

The above-mentioned ``decomposition" into collective and single particle components of the angular momentum is complex because it involves the transformation of the original basis states $P^I_{MK}\vert\kappa\rangle$ to the orthonormal basis states $\vert IMK\kappa\rangle$ and their mixing by the TPSM Hamiltonian. To arrive at a more intuitive picture, the particle and total angular momentum composition needs to be analyzed in a similar way as for the particle triaxial rotor model (see Ref. \cite{SSS} and earlier work cited therein). We intend to address this problem in a separate publication.

\begin{figure}[!ht]
		\begin{center}
			\hspace*{-0.2cm}\includegraphics[width=8.5cm, angle =0]{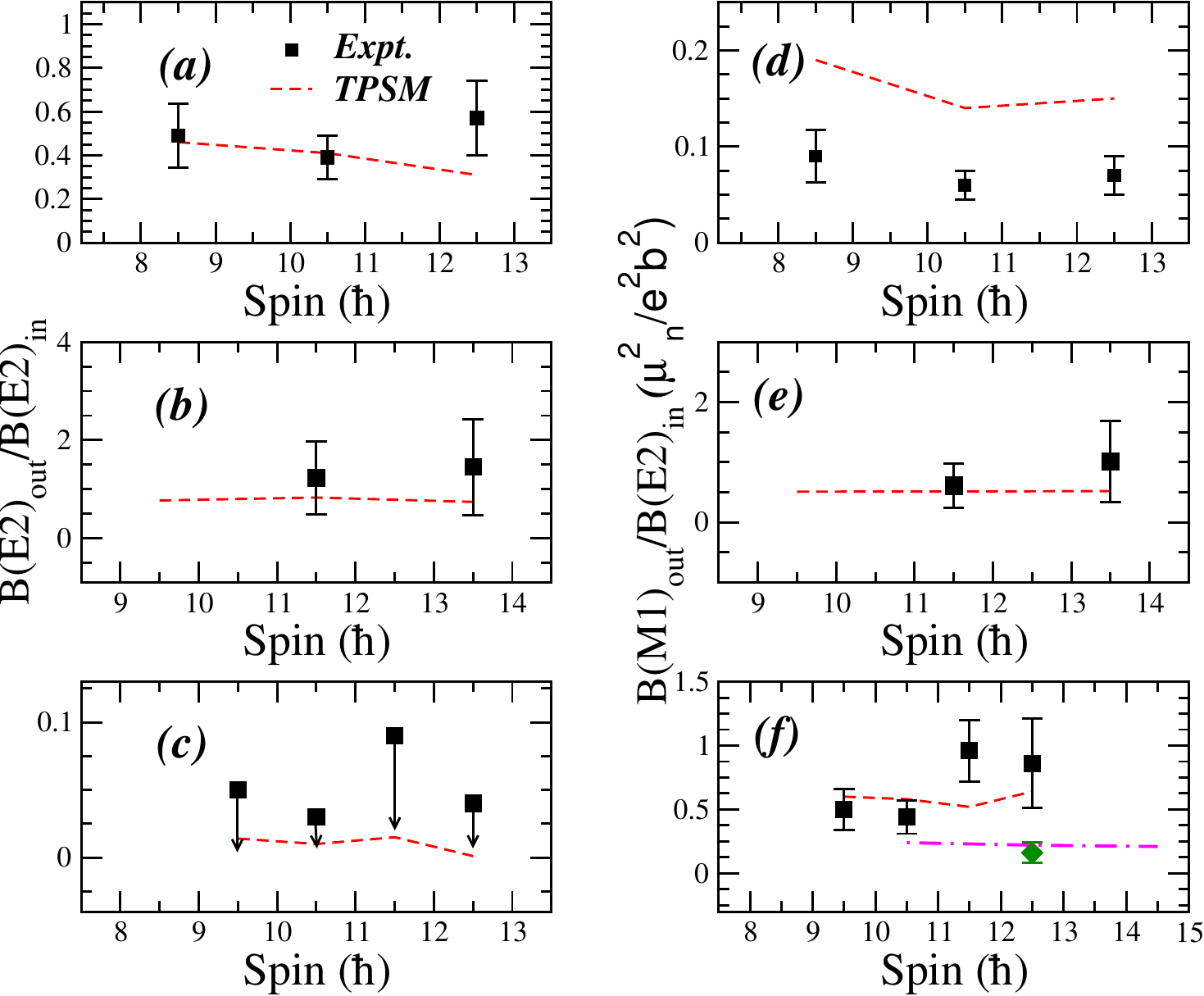}
		\end{center}
		\vspace{-0.5cm}
		\caption{The ratio of the rates of the out-band $\Delta$I = 1 transitions and the in-band E2 transitions of $^{105}$Pd are plotted as a function of spin, I. The E2 and M1 rates for the $\Delta$I = 1 transitions have been estimated using the evaluated values of the mixing ratios (Table~\ref{t1_pd}) and are shown on the left and right panels, respectively. The values for the Band 3 $\xrightarrow{}$ Band 1 transitions are shown in (a) and (d), the values for Band 4 $\xrightarrow{}$ Band 2 transitions in (b) and (e), and the values for the  Band 3 $\longleftrightarrow$ 4 transition in (c) and (f). Fig.~\ref{1_pd}(f) also includes the value for the Band 2 $\xrightarrow{}$ Band 1 transition (shown in green), assuming the E2 mixing to be negligible. In Fig.~\ref{1_pd}(c), the upper bounds correspond to the upper limit of the uncertainties of the mixing ratio values given in Table~\ref{t1_pd}. The dotted line in red on each panel represents the calculated values from TPSM, while the one in magenta in Fig.~\ref{1_pd}(f) corresponds to the calculated values for the Band 2 $\xrightarrow{}$ Band 1 transitions.}
		\label{1_pd}
	\end{figure}

\begin{figure}[!ht]
		\begin{center}
			\hspace*{-0.2cm}\includegraphics[width=8.5cm, angle =0]{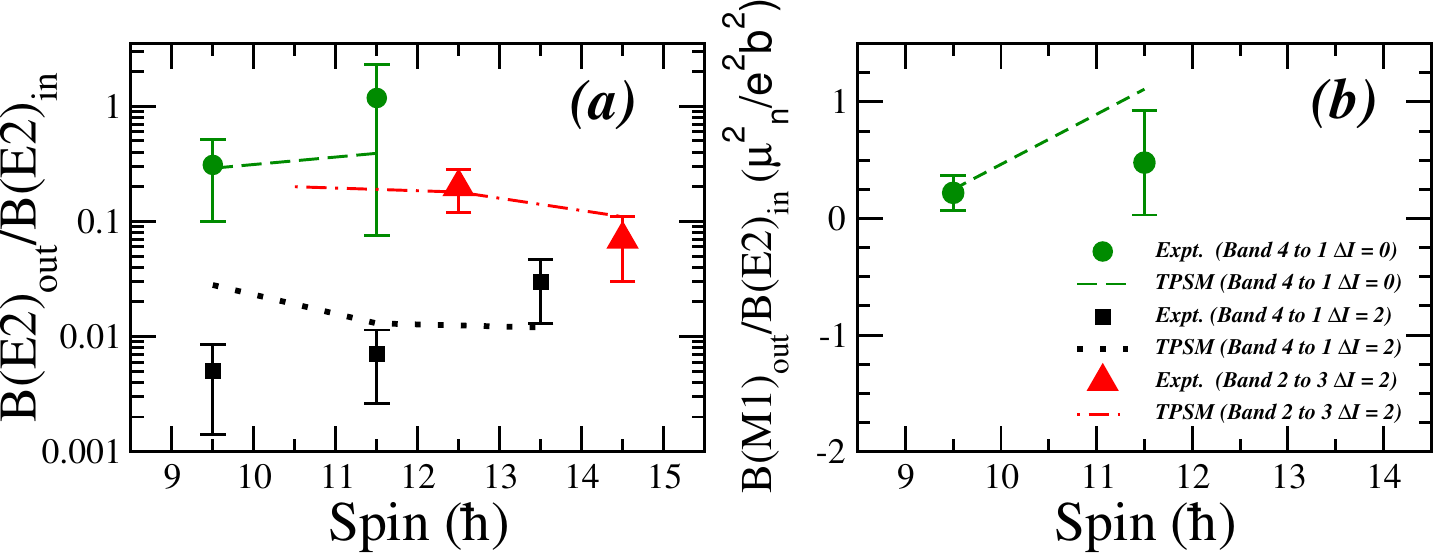}
		\end{center}
		\vspace{-0.5cm}
		\caption{The ratio of the rates of the out-band $\Delta$I = 0 and $\Delta$I = 2 transitions from Band 4 to Band 1, Band 2 to Band 3, and the in-band E2 transition of Band 4 as a function of spin, I. The M1 and E2 rates have been estimated using the evaluated values of the mixing ratios (Table.~\ref{t1_pd}) and plotted in (a) and (b), respectively. The dotted line on each panel represents the calculated values from TPSM.}
		\label{2_pd}
	\end{figure}

The TPSM wavefunctions have been used to calculate the M1 and E2 transition probabilities between the excited levels of $^{105}$Pd. The electric transitions have been evaluated with the effective charges of 1.6e for protons and 0.6e
for neutrons \cite{GH14, bh14, bh14a}. $g_l=0 (1.0)$ and $g_s=-3.826 (5.58)$ for neutrons (protons) effective charges have been used for M1 transitions.
The observed and calculated values of the ratio of the transition rates for the inter-band $\Delta I = 1$ transitions and intra-band E2 transition are listed in Table~\ref{t3_pd}. These values have also been plotted in Fig.~\ref{1_pd}. These ratios for the 
$\Delta I = 0$ and 2 inter-band transitions are plotted in Fig.~\ref{2_pd}.  The good agreement between the observed and calculated values substantiates the suggested interpretation.

 The values in Table~\ref{t3_pd} obey the following systematics. The $B(E2,\Delta I=1)$ probabilities for  Band 3 $\rightarrow$ Band 1 are enhanced, as expected for a collective transition connecting the $n=1$ TW state with the $n=0$ yrast state with the same quasineutron structure. The $B(M1,\Delta I=1)$ probabilities are moderate. In the same way the $B(E2,\Delta I=1)$ probabilities for the transitions, Band 2 $\rightarrow$ Band 4 are enhanced, as expected for a collective transition connecting the $n=1$ and the $n=0$ TW states with the same quasineutron structure. The $B(M1,\Delta I=1)$ probabilities are small.

  The $B(E2,\Delta I=1)$ probabilities for the transitions between Band 3 and Band 4, as well as Band 1 and Band 2, are weak because the $h_{11/2}$ quasineutron occupies different states with $k$ = 11/2 and 7/2, respectively. As only the quasineutron contributes, the $E2$ operator the $B(E2,\Delta I=1)$ is of single particle order. The $B(M1,\Delta I=1)$ values are also of single particle order, which is large for $h_{11/2}$ quasineutrons.
  
The $B(E2,\Delta I=0)$ values between Bands 1 and 4 and between Bands 2 and 3 are collectively enhanced, although not as much as the ones between the $n=0$ and 1 wobbling states.  The $B(M1,\Delta I=0)$ values are of the order that is typical for $h_{11/2}$ quasineutrons. As seen in Fig. \ref{wave}, the largest component of Bands 1, 3 is $\vert IM, K=11/2,\kappa=11/2,11/2\rangle$ and the largest component of Bands 1, 4 is $\vert IM, K=7/2,\kappa=11/2,11/2\rangle$, which only differ by a reorientation of $\bm{J}$ with respect to the short axis. As a consequence, the reduced matrix elements $\langle I,1\vert\vert E2\vert\vert I,4\rangle$ and  $\langle I,2\vert\vert E2\vert\vert I,3\rangle$ are enhanced. The contributions from the other components remain small because they add up in a non-coherent way. This is different for the reduced matrix elements $\langle I,1,2,3,4\vert\vert E2\vert\vert I-1,1,2,3,4\rangle$, where the contributions from the remaining components add up in a coherent way, such that the above-given interpretation of the $B(E2,\Delta I=1)$ values applies. 
  
Semiclassically, the enhanced $\Delta I=1$ transitions are generated by the wobbling motion of the whole charge distribution of the nucleus with a fixed orientation of the $h_{11/2}$ quasineutron with respect to the short axis. The enhanced $\Delta I=0$ transitions are generated by an oscillation between the two orientations of the whole charge distribution of the nucleus with respect to the angular momentum vector in the laboratory frame,  where the presence of an extra wobbling quanta does not matter. A more detailed analysis of the difference between the $\Delta I=1$ and $\Delta I=0$ transition matrix elements will be carried out separately. As seen in Figs. \ref{1_pd} and \ref{2_pd}, the measured branching ratios confirm the interpretation. Only the predicted $\Delta I=1$ transitions from Band 2 to 4 have not been observed in the present data. This may be understood from the fact that Band 2 has the least intensity, and these unobserved feed-out transitions are predominantly E2 in character, which need to compete with two other higher energy feed-out transitions - one M1 and the other E2. See Appendix \ref{secB} for an estimate.\par
 
The above interpretation in terms of the leading basis states loses relevance with increasing $I$ because of the mixing of different $K$ states projected from the quasineutron configuration at 1.31 MeV, as well as from the configurations at 1.07 MeV and 1.79 MeV change with $I$. This is reflected in the decrease in the probability of the configuration at 1.31 MeV with increasing $I$. Based on the detailed studies of $^{135}$Pr in the framework of the PTR model \cite{SSS, SCS}, we expect that the interpretation in terms of the different orientations of the $h_{11/2}$ quasineutron angular momentum remains qualitatively correct. 

 Our interpretation is at variance with the one suggested in Ref. \cite{105pd_prl}, where, based on the PTR model, the authors suggested that Band 2 is the signature partner of Band 1. According to this interpretation, it should originate from the quasineutron configurations with opposite signature, $k=9/2$. As seen in Fig. \ref{wave}, the $I=13/2$ state of Band 2 has it as the main component but changes to the discussed $k=7/2$ structure. The change is expected from Fig.~\ref{Band_before}. The lowest basis state with $k=9/2$ signature  $[1\nu,~13/2,~1.31]$ has about the same energy as $[1\nu,~7/2,~1.31]$ for $I=13/2$ but moves up for larger  $I$. This is reflected in the change in the structure of Band 2 between $I=13/2$ and 17/2 in Fig.~\ref{wave}. It also explains why the PTR calculations in  Ref. \cite{105pd_prl}, assuming a $9/2+2n$ signature, give too high energy for Band 2.
 
The nucleus $^{105}$Pd has a structure that is analogous to  $^{135}$Pr, where the odd $h_{11/2}$ quasineutron replaces the odd quasiproton. The latter has been studied in Refs. \cite{dona,garg,sensharmaa} using the TPSM and the PTR model. Both approaches predict that the second band with the signature  11/2, 15/2, 19/2, ... has the character of the $n$ = 2 double TW excitation mode, which is in agreement with the strong E2 transition to the $n$ = 1 single TW band observed in $^{135}$Pr \cite{sensharmaa}. As shown in the present work, this band has a different nature in $^{105}$Pd. The TPSM accounts for the different nature of these bands in  $^{105}$Pd and  $^{135}$Pr. At variance, the PTR  model predicts a $n$ = 2 double TW excitations for Band 4 in $^{105}$Pd as well \cite{Chen_PC}. The PTR model predicts Band 2 and the corresponding band in $^{135}$Pr at twice the energy above the yrast band than observed \cite{105pd_prl,sensharmaa}, while the TPSM approach reproduces the experimental energy difference quite well. The TPSM takes the anti-symmetrization between the odd nucleon and the nucleons of the collective rotor core into account, while the PTR model neglects it. This could be the reason for the discrepancy between the two models.\par

 \section{Summary}
 The multipolarities and electromagnetic character of the gamma transitions from the first excited band with $I=$13/2, 17/2, ... (Band 3) to the yrast sequence 11/2, 15/2, ... of Ref. \cite{105pd_prl} have been confirmed. The multipolarities and the electromagnetic characters of the gamma transitions from the excited levels of the second band with angular momentum states, 15/2, 19/2, 23/2, ... (Band 4) detected in Ref. \cite{pd_conf} have been determined through the R$\mathrm{_{DCO}}$ and polarization measurements. The transitions connecting it with the previously known single transverse wobbling band (Band 3) are dominated by the M1 component with only weak E2 admixtures. This excludes the interpretation as a double transverse wobbling excitation, which has been suggested for the analog band in $^{135}$Pr. Instead, Band 4 has been interpreted as another one-quasineutron band with the $h_{11/2}$ quasineutron having the reduced angular momentum projection $j_s=7/2$ on the short axis and the next higher band with $I=13/2$, 17/2, ... (Band 2) as transverse wobbling excitation built on Band 4. Consequently, the inter-band $\Delta I = 1$ transitions show a large enhancement in the B(E2) rates. Triaxial projected shell model calculations account well for the observed energies and ratios of the transition probabilities. Analyzing the calculated eigenstates demonstrated the nature of Band 3 as a wobbling excitation of Band 1, the nature of Band 2 as a wobbling excitation of Band 4, and the nature of Band 4 as another $h_{11/2}$ one-quasineutron configuration. However, it may be noted that the complete experimental validation of this interpretation needs future experiments with higher statistics, which should lead to the observation of the $\Delta I = 1$ transitions from Band 2 to 4 and allow the polarization measurements of the 254 and 300 keV transitions from Band 4 to 2.

In comparison to axial nuclei, triaxial systems have the possibility to rotate about the third axis as an additional collective degree of freedom, which generates the wobbling mode as a set of collective excitations. In general, the collective wobbling excitations are expected to appear not only based on the yrast band but also on quasiparticle excitations. The present work identified the first example of such a wobbling mode of a quasineutron excitation with a transverse wobbling excitation built on it, which is an essential indication of the collective nature of the wobbling mode.

\appendix
\section{Change of the quantization axis}\label{secA}
 Technically, it is achieved by changing the triaxiality parameter $\gamma$ to an equivalent value in another sector. The TPSM parameters $\epsilon=0.26$ and $\epsilon'=0.11$ correspond to $\epsilon=\epsilon_0\cos\gamma$ and $\epsilon'=\epsilon_0\sin\gamma$ with $\epsilon_0=0.28$ and $\gamma=23^\circ$ and the 3-axis being long principal axis. Changing $\gamma\rightarrow-120^\circ-\gamma=-143^\circ$ leaves the results invariant but moves the 3-axis to the short principal axis, and the TPSM calculations were performed for $\epsilon=0.22$ and $\epsilon'=-0.17$. 
 \section{Intensity estimate of the missing transitions from Band 2 to 4.}\label{secB}
 In the present data set, we could observe the 1158 keV M1 transition and 939 keV E2 transition from the 21/2$^-$ level of Band 2, but the competing transition probability of the 555 keV E2 transition to Band 4 becomes significantly smaller. This intensity can be estimated using the calculated B(M1) and B(E2) rates for the $I$ = 21/2$^-$ level of Band 2 as given in Table~\ref{t2_pd} and was found to be 0.10 or 0.22 when normalized with the intensity of 1158 keV or 939 keV, respectively. In addition, this small intensity of the 555 keV transition gets further divided in the four decay transitions from the 19/2$^-$ level of Band 4 in almost equal proportion (refer to the intensities given in Table~\ref{t1_pd}), which is not the case for the intensities of the 1158 and 939 keV transitions. Thus, in order to detect this transition, it is necessary to have a larger detector array, which will lead to substantial statistics in the 3-fold coincidence. In this case, it will be possible to generate a double-gated spectrum with one gate list consisting of the in-band transitions of Band 2 and the other gate list with the four feed-out transitions of the 19/2$^-$ level of Band 4.

\section*{Acknowledgements}
Private communications on the particle-rotor model results for $^{105}$Pd with Prof. Q. B. Chen are gratefully acknowledged. GHB, JAS, and NR acknowledge the Science and Engineering Research Board (SERB), Department of Science and Technology, Govt. of India, for providing financial support under Project No. CRG/2019/004960 to carry out a part of this research work. AK is thankful to the Council of Scientific $\&$ Industrial Research (CSIR), India, for the Senior Research Fellowship vide file no. 09/489(0121)/2019-EMR-I.

\bibliographystyle{apsrev4-1}

\begin{thebibliography}{0}%
\makeatletter
\providecommand \@ifxundefined [1]{%
 \@ifx{#1\undefined}
}%
\providecommand \@ifnum [1]{%
 \ifnum #1\expandafter \@firstoftwo
 \else \expandafter \@secondoftwo
 \fi
}%
\providecommand \@ifx [1]{%
 \ifx #1\expandafter \@firstoftwo
 \else \expandafter \@secondoftwo
 \fi
}%
\providecommand \natexlab [1]{#1}%
\providecommand \enquote  [1]{``#1''}%
\providecommand \bibnamefont  [1]{#1}%
\providecommand \bibfnamefont [1]{#1}%
\providecommand \citenamefont [1]{#1}%
\providecommand \href@noop [0]{\@secondoftwo}%
\providecommand \href [0]{\begingroup \@sanitize@url \@href}%
\providecommand \@href[1]{\@@startlink{#1}\@@href}%
\providecommand \@@href[1]{\endgroup#1\@@endlink}%
\providecommand \@sanitize@url [0]{\catcode `\\12\catcode `\$12\catcode
  `\&12\catcode `\#12\catcode `\^12\catcode `\_12\catcode `\%12\relax}%
\providecommand \@@startlink[1]{}%
\providecommand \@@endlink[0]{}%
\providecommand \url  [0]{\begingroup\@sanitize@url \@url }%
\providecommand \@url [1]{\endgroup\@href {#1}{\urlprefix }}%
\providecommand \urlprefix  [0]{URL }%
\providecommand \Eprint [0]{\href }%
\providecommand \doibase [0]{http://dx.doi.org/}%
\providecommand \selectlanguage [0]{\@gobble}%
\providecommand \bibinfo  [0]{\@secondoftwo}%
\providecommand \bibfield  [0]{\@secondoftwo}%
\providecommand \translation [1]{[#1]}%
\providecommand \BibitemOpen [0]{}%
\providecommand \bibitemStop [0]{}%
\providecommand \bibitemNoStop [0]{.\EOS\space}%
\providecommand \EOS [0]{\spacefactor3000\relax}%
\providecommand \BibitemShut  [1]{\csname bibitem#1\endcsname}%
\let\auto@bib@innerbib\@empty
\end{thebibliography}%


\begin{thebibliography}{99}

\bibitem{triaxial} P. M\"oller, R. Bengtsson B. G. Gillis, P. Olivius, T. Ichikawa, Phys. Rev. Lett. \textbf{97} 162502 (2006).
		
\bibitem{bohr}  A. Bohr and B. R. Mottelson, \textit{Nuclear Structure} (W. A. Benjamin, New York, 1975), Vol. II, Chap. 4.
		
\bibitem{prl_hageman}  S. W. \O{}deg\aa{}rd \textit{et al.} Phys. Rev. Lett \textbf{86}, 5866 (2001).
		
\bibitem{Bengtsson04} R. Bengtsson and H. Ryde, Eur. Phys. J. A {\bf 22}, 355 (2004).
		
\bibitem{dona} S. Frauendorf and F. D\"onau, Phys. Rev. C \textbf{89}, 014322 (2014).
		
\bibitem{garg} J. T. Matta \textit{et al.} Phys. Rev. Lett. \textbf{114}, 082501 (2015).
		
\bibitem{nandi} S. Nandi \textit{et al.} Phys. Rev. Lett. \textbf{125}, 132501 (2020).
		
		
\bibitem{127xe} S. Chakraborty \textit{et al.}Phys. Lett. B \textbf{811}, 135854 (2020).
		
		
\bibitem{151eu} A. Mukherjee \textit{et al.} Phys. Rev. C \textbf{107}, 054310 (2023).

\bibitem{163lu} D. R. Jensen \textit{et al.} Phys. Rev. Lett. \textbf{89}, 142503 (2002).
		
\bibitem{161lu} P. Bringel \textit{et al.} Eur. Phys. J. A \textbf{24}, 167 (2005).
		
\bibitem{165lu} G Sch\"onwa\ss er \textit{et al.} Phys. Lett. B \textbf{552}, 9 (2003).
		
\bibitem{167lu} H Amro \textit{et al.} Phys. Lett. B \textbf{553}, 197 (2003).
		
\bibitem{167ta} D. J. Hartley \textit{et al.} Phys. Rev. C \textbf{80}, 041304(R) (2009).
		
\bibitem{133la} S. Biswas \textit{et al.} Eur. Phys. J. A \textbf{55}, 159 (2019).
		
\bibitem{135pr} Kosai Tanabe and Kazuko Sugawara-Tanabe Phys. Rev. C \textbf{95}, 064315 (2017).
		
\bibitem{sensharmaa} N. Sensharma \textit{et al.} Phys. Lett. B {\bf 792} 170 (2019).
		
\bibitem{sensharmaa2} N. Sensharma \textit{et al.}  Phys. Rev. Lett. \textbf{124}, 052501 (2020).

\bibitem{105pd_prl} J. Tim\'ar \textit{et al.} Phys. Rev. Lett. \textbf{122}, 062501 (2019).

\bibitem{pd_old} F. A. Rickey, J. A. Grau, L. E. Samuelson and P. C. Simms, Phys. Rev. C  \textbf{15}, 1530 (1977).
		
		
\bibitem{pd_conf} J. Tim\'ar \textit{et al.} J. Phys.: Conf. Ser. \textbf{1555} 012025 (2020).

		

\bibitem{marcos} R. Palit, S. Saha, J. Sethi, T. Trivedi, S. Sharma, B. S. Naidu, S. Jadhav, R. Donthi, P. B. Chavan, H. Tan and W. Hennig, Nucl. Inst. Methods A \textbf{680}, 90 (2012).

\bibitem{rad} D. C. Radford, Nucl. Inst. Methods A \textbf{361}, 297 (1995).

\bibitem{iaeaINISRepository} R. Bhowmik, S. Muralithar, R.P. Singh Proc. DAE Symp. Nucl. Phys. \textbf{44 B} 422 (2001).
		
\bibitem{dco}  A. Kr\"{a}mer-Flecken \textit{et al.} Nuclear Instruments and Methods in Physics Research Section A: Accelerators, Spectrometers, Detectors and Associated Equipment \textbf{275}, 333 (1989).
		
\bibitem{105pd_ang} E. S. Macias, W. D. Ruhter, D. C. Camp and R. G. Lanier Comp. Phys. Comm. \textbf{11}, 75 (1976).
        

		
		
\bibitem{rev_tpsm} K. Hara and Y. Sun, Int. J. Mod. Phys. E {\bf 4}, 637 (1995).
        
\bibitem{tpsm}  J. A. Sheikh and K. Hara, Phys. Rev. Lett. {\bf 82}, 3968 (1999).

\bibitem{rev_tpsm_tpsm} Javid A Sheikh, Gowhar H Bhat, Waheed A Dar, Sheikh Jehangir and Prince A Ganai, Phys. Scr. {\bf 91},   063015  (2016).
		
\bibitem{SJ21} S. Jehangir, G. H. Bhat, N. Rather, J. A. Sheikh and R. Palit, Phys. Rev. {\bf C 104}, 044322 (2021).
		
\bibitem{SB10} J. A. Sheikh, G. H. Bhat, Y. Sun and R. Palit, Phys. Lett. {\bf B 688}, 305 (2010).
		
\bibitem{JM17} J. Marcellino \textit{et al.}  Phys. Rev. {\bf C 96}, 034319 (2017).

\bibitem{https://doi.org/10.1016/0375-9474(68)90044-4} Michel Baranger and Krishna Kumar,  Nucl. Phys. A {\bf 122},  241-272 (1986).

\bibitem{Hara332} K. Hara and S. Iwasaki, Nucl. Phys. A \textbf{332}  (1979) 61.

\bibitem{Hara348} K. Hara and S. Iwasaki, Nucl. Phys. A \textbf{348} (1980) 200.
\bibitem{ring80}P. Ring and P. Schuck, {\it The Nuclear Many Body Problem}(Springer-Verlag, New York), (1980).
  
\bibitem{45jm} W. Dieterich, A. Bäcklin, C.O. Lannergård, I. Ragnarsson, Nucl. Phys. {\bf A 253} (1975) 429.
  
\bibitem{Rouoof25}S. P. Rouoof, Nazira Nazir, S. Jehangir, G. H. Bhat, J. A. Sheikh, N. Rather, and S. Frauendorf  Phys. Rev. C \textbf{111}, 054309 (2025).

\bibitem{GH14} G. H. Bhat, W. A. Dar, J. A. Sheikh and Y. Sun, Phys. Rev. C {\bf  89}, 014328 (2014).
		
\bibitem{bh14} G. H. Bhat, J. A. Sheikh W. A. Dar, S. Jehangir, R. Palit and P. A. Ganai,   Phys. Lett. B {\bf 738}, 218 (2014).
		
\bibitem{bh14a} G. H. Bhat, R. N. Ali, J. A. Sheikh and  R. Palit, Nucl. Phys. A {\bf  922}, 150  (2014).

\bibitem{wang20} L. J. Wang, F. Q. Chen, and Y. Sun, Phys. Lett. B \textbf{808}, 135676 (2020).

\bibitem{nazir23} N. Nazir, S. Jehangir, S. P. Rouoof, G. H. Bhat, J. A. Sheikh, N. Rather, and M. A. Malik, Phys. Rev. C \textbf{108}, 044308 (2023).

\bibitem{SSS} Q. B. Chen and S. Frauendorf, Phys. Rev. C {\bf 109}, 044304 (2024).

\bibitem{SCS} Q. B. Chen and S. Frauendorf, Eur. Phys. J. A {\bf 58}, 75 (2022).

\bibitem{Chen_PC} Q. B. Chen Private Communication.












\end{thebibliography}

\end{document}